\documentclass[
  aps,prd,twocolumn,superscriptaddress,
  nofootinbib,longbibliography,floatfix
]{revtex4-2}

\usepackage{amsmath,amssymb,mathtools}
\usepackage{bm}
\usepackage{graphicx}
\usepackage{xcolor}
\usepackage[colorlinks=true,allcolors=blue]{hyperref}

\newcommand{\dCS}{dynamical Chern--Simons}

\newcommand{\rs}{r_{*}}
\newcommand{\Li}{\mathrm{Li}}
\newcommand{\zzeta}{\zeta}                 

\newcommand{\Ord}{\mathcal{O}}
\newcommand{\Voff}{V_{\mathrm{off}}}
\newcommand{\Gax}{\Gamma_{\mathrm{ax}}}
\newcommand{\Gpol}{\Gamma_{\mathrm{pol}}}
\newcommand{\Ggs}{\Gamma_{g\to s}}

\begin{document}

\title{Analytic parity-violating greybody factors of a dynamical Chern–Simons black hole}

\author{Abhishek Rout}
\email{arout@email.sc.edu}
\affiliation{
  Department of Physics and Astronomy,\\
  University of South Carolina, Columbia, South Carolina 29208, USA
}

\date{\today}

\begin{abstract}
We study the low-frequency scattering of gravitational and
pseudoscalar waves off a non-rotating black hole in \dCS{} (dCS)
gravity, where the axial (odd-parity) gravitational sector is
irreducibly coupled to the dynamical pseudoscalar while the polar
sector remains general relativistic. Working to first order in the
dimensionless coupling $\zzeta=(\ell_{\rm CS}/M)^4$ and using matched
asymptotic expansions, we obtain, for the quadrupole ($\ell=2$)
channel, an exact closed form for the graviton$\to$pseudoscalar
conversion probability and a high-precision analytic reconstruction of
the absorption (greybody) asymmetry between the axial and polar
sectors. Both reduce to combinations of $\pi^4$, $\pi^2$, and rational
numbers, with no $\ln2$ or odd-zeta terms in the reconstructed basis;
by contrast, the transmission-phase sector is governed by an
analytically derived static integral involving Ap\'ery's constant
$\zeta(3)$, with its overall matching prefactor identified numerically.
The parity-violating reflection-phase asymmetry is strongly suppressed
at low frequency---well below the parity-even general-relativistic
reflection phase---so that any polarization rotation imprinted on the
reflected wave is negligible, while a subleading $\Ord(\omega\zzeta)$
asymmetry remains in the transmitted phase. These effects are
intrinsically strong-field and low-frequency: the normalized flux
asymmetries are numerically consistent with a $1/(M\omega)^2$ falloff
in the geometric-optics limit, where the leading remaining flux effect
is conversion loss. The matched-asymptotic results are checked against
direct two-channel integration, and the reduced one-dimensional
integral determining the absorption asymmetry is evaluated to high
precision within an analytically controlled transcendental basis. The
coefficients provide analytic targets linking dCS parity violation in
absorption and mode conversion to the axial/polar quasinormal-mode
splitting.
\end{abstract}

\maketitle

\section{Introduction}
\label{sec:intro}

Parity is not a symmetry we are entitled to assume in gravity. A
parity-violating (PV) sector in the gravitational interaction would
leave distinctive fingerprints on the polarization of gravitational
waves (GWs)---most notably an amplitude and phase birefringence
between left- and right-handed states---and searching for such
imprints has become a concrete goal of GW astronomy
\cite{LueWangKamionkowski1999,AlexanderPeskin2006,%
AlexanderYunes2009review,Yunes2016implications,Okounkova2022birefringence,LVKtestingGR}. Among the
candidate theories, \dCS{} (dCS) gravity occupies a
privileged position: augmenting general relativity (GR) by the canonical
four-derivative, parity-odd curvature invariant---the Pontryagin
density---coupled to a dynamical pseudoscalar field, it arises in string theory, in loop quantum gravity, and in
effective-field-theory completions of gravity, and it serves as the
canonical benchmark for strong-field parity violation
\cite{JackiwPi2003,AlexanderYunes2009review,%
Smith2008,AlexanderYunes2007PRL}.

Because the background Pontryagin density vanishes for any
spherically symmetric spacetime, the Schwarzschild solution persists
in \dCS{} gravity, and the leading signatures appear at the level of
perturbations \cite{YunesPretorius2009,KonnoMatsuyamaTanda2009}. The
key structural fact is that the pseudoscalar couples only to the
odd-parity (axial) gravitational sector, leaving the even-parity
(polar) sector governed by the Zerilli equation and hence identical
to GR. Since the Regge--Wheeler and Zerilli potentials are
isospectral in GR \cite{Chandrasekhar1983}, the axial--polar
degeneracy of both the quasinormal-mode (QNM) spectrum and the
transmission (greybody) factors is a clean null of GR that \dCS{}
gravity breaks at first order in the coupling~\cite{LiIsospectrality2024}. The QNM side of this
isospectrality breaking has been studied in detail: the coupled
axial--pseudoscalar system was derived and analyzed by Cardoso and
Gualtieri \cite{CardosoGualtieri2009} and by Molina \emph{et al.}
\cite{Molina2010}, and the modified ringdown spectrum has since been
computed to increasing accuracy \cite{WagleYunesSilva2021,ChungLamYunes2025,HuLanGuo2026}, using slow-rotation QNM techniques developed in parallel for related higher-curvature theories such as Einstein-dilaton–Gauss-Bonnet gravity \cite{PieriniGualtieri2021,PieriniGualtieri2022}. Observational bounds on the \dCS{} length scale
now follow from a combination of GW and neutron-star data
\cite{Nair2019,SilvaNICER2021,AlexanderYunes2009review}.

What has been missing is the \emph{scattering} counterpart of this
picture. The greybody factors---the frequency-dependent transmission
probabilities that govern absorption, Hawking emission, and
wave scattering off the barrier
\cite{StarobinskyChurilov1973,Page1976,DasGibbonsMathur1997,Kanti2004}---encode the same
isospectrality breaking as the QNMs, but in a channel that also
contains a process the ringdown of the metric sector does not: the direct conversion of an axial gravitational perturbation into
propagating \dCS{} pseudoscalar radiation.
To date these transmission-sector observables of \dCS{} gravity have
not been given in closed form, and it is not obvious \emph{a priori}
that they should admit one.

Phenomenologically, these transmission coefficients are the building
blocks of larger observables: summed over multipoles they give the
parity-dependent absorption cross section and the modified Hawking
spectrum, while the conversion probability fixes the fraction of
incident gravitational flux radiated into the pseudoscalar. Two
features distinguish the \dCS{} case from parity-even higher-curvature
theories. First, because the pseudoscalar couples only to the axial
sector, the leading effect is a clean \emph{asymmetry} between the
parity channels---the polar sector remaining general-relativistic---%
rather than a shift of both parity sectors, as occurs when the extra
field couples to even parity as well (e.g.\ in
Einstein-dilaton--Gauss-Bonnet gravity). Second, the coupled system
supports a genuinely new \emph{conversion} channel, absent from the
single-field metric ringdown, in which an axial gravitational
perturbation leaks energy into the \dCS{} scalar. In the static
($\omega\to0$) limit the same near-zone problem underlies the tidal
response of the hole, so the coefficients obtained here are the
dissipative counterparts of the conservative tidal Love numbers of
non-spinning \dCS{} black holes~\cite{CardosoFranzin2017}, and connect
to the effective-field-theory matching programme for the
theory~\cite{CassemHertzberg2026}. They thus provide
normalization-fixed analytic benchmarks, complementary to the
ringdown, for any numerical-relativity or data-analysis study of dCS
parity violation.

In this paper we obtain an exact closed form for the conversion probability and a high-precision reconstruction of the absorption asymmetry. Working to first order in the dimensionless coupling $\zeta=(\ell_{\rm CS}/M)^4$ and to leading order in $M\omega$, we solve the coupled axial–pseudoscalar system of a Schwarzschild dynamical Chern–Simons black hole by matched asymptotic expansions. For the quadrupole channel, the conversion coefficient is evaluated analytically, while the absorption coefficient is reduced to a one-dimensional static integral and reconstructed within an analytically constrained transcendental basis: the graviton$\to$pseudoscalar
conversion probability
\begin{equation}
d_2=\frac{3(\pi^2-9)^2}{8},
\end{equation}
and the graviton greybody (absorption) asymmetry
\begin{equation}
\kappa_2=\frac{21}{80}\pi^4+\frac{27}{8}\pi^2-\frac{89833}{1536}.
\end{equation}
Both reduce to combinations of $\pi^4$, $\pi^2$, and rational numbers, with no $\ln{2}$ or odd-zeta terms in the reconstructed basis. The \emph{dispersive} phase sector, by contrast, is governed by Ap\'ery's constant $\zeta(3)$, while the parity-violating reflection-phase asymmetry is strongly
suppressed at low frequency---well below the parity-even
general-relativistic reflection phase---so that the reflected wave
suffers negligible polarization rotation (a subleading $\Ord(\omega\zzeta)$ asymmetry remains in the transmitted phase). We further show, through the geometric-optics limit, that the normalized flux asymmetries are intrinsically strong-field and low-frequency, with a high-frequency falloff numerically consistent with $(M\omega)^{-2}$, where only conversion loss survives. The matched-asymptotic results are checked against direct two-channel integration at accessible low-frequency precision, while the reduced integral for $\kappa_{2}$ is independently evaluated to high precision and reconstructed within the analytically controlled basis.

Beyond their intrinsic interest, these results provide analytic targets---normalization-independent once the scalar normalization and the definition of $\zzeta$ are fixed--- that any numerical-relativity study or data-analysis pipeline probing \dCS{} parity violation can calibrate
against, and they are complementary to the ringdown: the absorption
asymmetry is correlated with the axial/polar QNM splitting, while the
conversion channel is unique to scattering. The strong-field,
low-frequency character of the effect situates it in the regime
relevant to ringdown and to space-based observations of massive black
holes \cite{Berti2015review,BertiSpectroscopy2026,LISA2017,BarausseEtAl2020}, and the
pseudoscalar-emission channel connects naturally to light dark-sector phenomenology \cite{NgIsiHaster,BritoCardosoPani2015} once a mass is restored.

The paper is organized as follows.
Section~\ref{sec:framework} sets up \dCS{} gravity and the coupled
axial--scalar master equations.
Section~\ref{sec:mae} develops the low-frequency matched-asymptotic
framework and the static solutions.
Section~\ref{sec:results} derives the coefficients
$d_2$ and $\kappa_2$ and discusses their transcendental structure.
Section~\ref{sec:phys} interprets the physical observables, including the suppressed reflection-phase birefringence and the absence of net Hawking circular polarization.
Section~\ref{sec:checks} collects limits, checks, and the multipole
dependence, and Section~\ref{sec:discussion} discusses the
observational context and extensions. Technical details are relegated
to the appendices. We use geometric units $G=c=1$ and metric
signature $(-,+,+,+)$.

\section{Theoretical framework}
\label{sec:framework}

\subsection{Action and field equations}
\label{sec:action}

We work in \dCS{} gravity with a canonical, massless
pseudoscalar field $\vartheta$. In geometric units $G=c=1$ and with
metric signature $(-,+,+,+)$, the action is
\begin{align}
S =\int d^4x\,\sqrt{-g}\,\Big[
  &\,\kappa\,R
  +\frac{\alpha}{4}\,\vartheta\,{}^{*}\!RR
  \nonumber\\
  &-\frac{\beta}{2}\big(\nabla_a\vartheta\,\nabla^a\vartheta
     +2V(\vartheta)\big)\Big],
\label{eq:action}
\end{align}
where $\kappa=(16\pi)^{-1}$, and
${}^{*}\!RR\equiv
{}^{*}\!R^{a}{}_{b}{}^{cd}R^{b}{}_{acd}$
is the Pontryagin density, built from the dual Riemann tensor
${}^{*}\!R^{a}{}_{b}{}^{cd}=\tfrac12\,
\varepsilon^{cdef}R^{a}{}_{bef}$.
The constants $\alpha$ and $\beta$ set, respectively, the strength of
the parity-violating coupling and the pseudoscalar kinetic term. We
take the baseline massless case $V=0$; a mass term (the axion-like /
dark-matter extension) is discussed in
Sec.~\ref{sec:discussion}.

Varying Eq.~\eqref{eq:action} with respect to the metric and the
scalar gives
\begin{align}
R_{ab} &= -16\pi\alpha\,C_{ab}
          +8\pi\beta\,\nabla_a\vartheta\,\nabla_b\vartheta ,
\label{eq:einstein}\\[2pt]
\beta\,\Box\vartheta &= -\frac{\alpha}{4}\,{}^{*}\!RR ,
\label{eq:scalareom}
\end{align}
where the C-tensor
\begin{equation}
C^{ab}=\nabla_c\vartheta\,
        \varepsilon^{cde(a}\nabla_e R^{b)}{}_{d}
       +\nabla_c\nabla_d\vartheta\,
        {}^{*}\!R^{d(ab)c}
\label{eq:ctensor}
\end{equation}
carries the parity violation. Equation~\eqref{eq:scalareom} shows
that a nonvanishing Pontryagin density sources the pseudoscalar,
which in turn back-reacts on the axial gravitational sector through
Eq.~\eqref{eq:einstein}---the mechanism responsible for every effect
in this paper.

\subsection{Background and dimensionless coupling}
\label{sec:background}

The Schwarzschild metric with $f(r)=1-2M/r$,
\begin{equation}
ds^2=-f\,dt^2+f^{-1}dr^2+r^2 d\Omega^2 ,
\end{equation}
together with $\bar\vartheta=0$ is an exact solution of
Eqs.~\eqref{eq:einstein}--\eqref{eq:scalareom}, because the
background Pontryagin density vanishes,
${}^{*}\!\bar R\bar R=0$. The parity-violating physics therefore
appears only at the level of \emph{perturbations}, through the
linearized Pontryagin source $\delta({}^{*}\!RR)\neq0$. Since the
background carries no scalar hair, the leading effects are
spin-independent and set entirely by the barrier region.

Following the normalization of
Refs.~\cite{CardosoGualtieri2009,Molina2010} we fix $\alpha=1$ and
measure the coupling through the single dimensionless combination
\begin{equation}
\zzeta \equiv \frac{16\pi}{\beta M^4}
      = \left(\frac{\ell_{\rm CS}}{M}\right)^{\!4},
\label{eq:zeta}
\end{equation}
where $\ell_{\rm CS}$ is the \dCS{} length scale. All results below
are obtained to first order in $\zzeta$. The effective-field-theory
treatment of \dCS{} gravity may apply for $\zzeta$ below order unity,
while the perturbative expansion employed here requires $\zzeta\ll1$.

In the conventions of Refs.~\cite{CardosoGualtieri2009,Molina2010} the
pseudoscalar $\vartheta$ and the kinetic coupling $\beta$ are
dimensionless, while $\alpha$ has dimensions of (length)$^2$, so that
both $\alpha\,\vartheta\,{}^{*}\!RR$ and $\beta(\nabla\vartheta)^2$ scale
as the Einstein--Hilbert term with $\kappa=(16\pi)^{-1}$. The \dCS{}
length scale is defined by
$\ell_{\rm CS}^4\equiv\alpha^2/(\kappa\beta)=16\pi\alpha^2/\beta$,
which for $\alpha=1$ reduces to Eq.~\eqref{eq:zeta}. A healthy scalar
kinetic term requires $\beta>0$, hence $\zzeta>0$. Other common
normalizations rescale $\vartheta$, $\alpha$, and $\beta$, and thereby
the numerical value of $\ell_{\rm CS}$ and $\zzeta$; all coefficients
quoted here refer to the convention of Eq.~\eqref{eq:zeta}.

\subsection{Coupled axial--scalar master equations}
\label{sec:master}

Decomposing the metric perturbation into odd- and even-parity
tensor harmonics and the scalar into
$\delta\vartheta=\Theta(r)\,Y_{\ell m}/r\,e^{-i\omega t}$, one finds
that the even (polar) gravitational sector decouples from
$\vartheta$ and obeys the Zerilli equation---it is therefore
\emph{identical to general relativity}. The odd (axial) sector, by
contrast, mixes the Regge--Wheeler master variable $\Psi$ with the
pseudoscalar. In the tortoise coordinate
$\rs=r+2M\ln(r/2M-1)$ the system takes the two-channel form
\begin{equation}
\left[\frac{d^2}{d\rs^2}+\omega^2\right]\!
\begin{pmatrix}\Psi\\[2pt]\Theta\end{pmatrix}
=
\begin{pmatrix}V_{11}&\Voff\\[2pt]\Voff&V_{22}\end{pmatrix}\!
\begin{pmatrix}\Psi\\[2pt]\Theta\end{pmatrix},
\label{eq:coupled}
\end{equation}
with, after symmetrization,
\begin{subequations}
\label{eq:potentials}
\begin{align}
V_{11}&=f\!\left[\frac{\ell(\ell+1)}{r^2}-\frac{6M}{r^3}\right],
\\[2pt]
V_{22}&=f\!\left[\frac{\ell(\ell+1)}{r^2}
        \Big(1+36\,\zzeta\,\tfrac{M^6}{r^6}\Big)
        +\frac{2M}{r^3}\right],
\\[2pt]
\Voff&=6\sqrt{\Lambda}\,\sqrt{\zzeta}\;\frac{M^3 f}{r^5},
\qquad
\Lambda=\frac{(\ell+2)!}{(\ell-2)!}.
\label{eq:Voff}
\end{align}
\end{subequations}

\begin{figure}[t]
\centering
\includegraphics[width=\columnwidth]{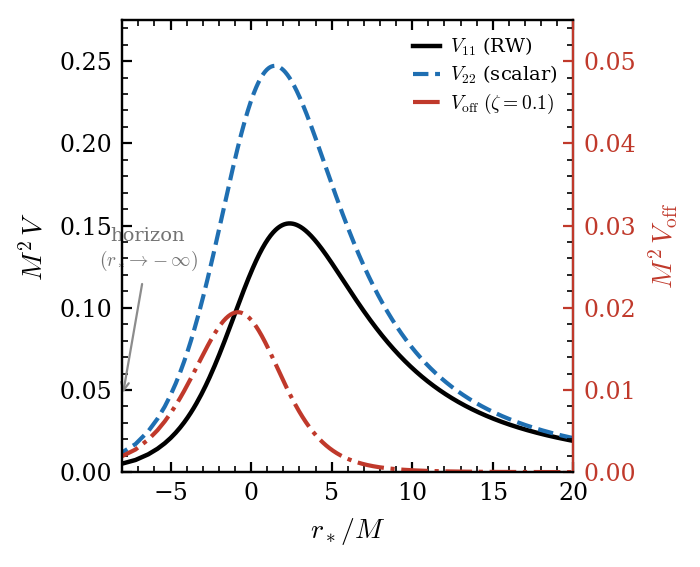}
\caption{Effective potentials of the coupled $\ell=2$ system versus
tortoise coordinate ($M=1$): the Regge--Wheeler barrier $V_{11}$
(axial graviton), the scalar barrier $V_{22}$, and the off-diagonal
coupling $\Voff$ (right axis, $\zzeta=0.1$). The diagonal barriers
differ by the spin-2/spin-0 gap $V_{11}-V_{22}=-8Mf/r^3<0$ and cannot
be removed by a constant rotation, so the system is irreducibly
two-channel; the coupling is localized in the strong-field region
near the peaks and is parametrically smaller, $\Ord(\sqrt{\zzeta})$.
As $\zzeta\to0$ the channels decouple into the Regge--Wheeler and
scalar barriers.}
\label{fig:potentials}
\end{figure}

Here $V_{11}$ is the spin-2 Regge--Wheeler potential and
$V_{22}|_{\zzeta=0}$ the spin-0 potential; we have verified
Eqs.~\eqref{eq:coupled}--\eqref{eq:potentials} against the coupled
system of Refs.~\cite{CardosoGualtieri2009,Molina2010}
(see Appendix~\ref{app:master}).

Two structural features control everything that follows. First, the
diagonal potentials differ by the spin-2/spin-0 gap
$V_{11}-V_{22}=-8Mf/r^3\neq0$, so the channels cannot be decoupled by
a constant rotation: the system is irreducibly two-channel. Second,
the coupling is off-diagonal and $\Ord(\sqrt{\zzeta})$; the
graviton$\to$scalar conversion \emph{amplitude} is therefore
$\Ord(\sqrt{\zzeta})$ and its \emph{probability} $\Ord(\zzeta)$,
the same order at which the axial and polar greybody factors first
differ. At $\zzeta\to0$ the axial channel reduces to Regge--Wheeler,
which is isospectral to the polar Zerilli channel, so every
parity-violating observable defined below is a pure $\Ord(\zzeta)$
quantity (Fig.~\ref{fig:potentials}).

\section{Low-frequency matched asymptotics}
\label{sec:mae}

\subsection{Scattering observables}
\label{sec:observables}

We consider an axial gravitational wave of frequency $\omega$
incident from infinity on the black hole. With purely ingoing
conditions at the horizon, the two-channel solution of
Eq.~\eqref{eq:coupled} behaves as
\begin{align}
\rs\to+\infty:\quad
&\Psi\sim e^{-i\omega\rs}+R_{11}e^{+i\omega\rs},
&&\Theta\sim R_{21}e^{+i\omega\rs},
\nonumber\\
\rs\to-\infty:\quad
&\Psi\sim T_{11}e^{-i\omega\rs},
&&\Theta\sim T_{21}e^{-i\omega\rs},
\end{align}
and flux conservation reads
$|R_{11}|^2+|R_{21}|^2+|T_{11}|^2+|T_{21}|^2=1$. We define the
axial-graviton greybody factor, the graviton$\to$scalar conversion
probability, and (as the parity-even reference) the polar greybody
factor,
\begin{equation}
\Gax=|T_{11}|^2,\qquad
\Ggs=|T_{21}|^2,\qquad
\Gpol=\Gamma^{\rm GR}_{\rm RW},
\end{equation}
where $\Gpol$ is the Regge--Wheeler value, equal by isospectrality to
the polar Zerilli greybody factor. Since $\Gax=\Gpol$ at
$\zzeta=0$, the two leading parity-violating observables are pure
$\Ord(\zzeta)$ ratios,
\begin{align}
d_\ell &\equiv \frac{\Ggs}{\zzeta\,\Gpol}
   \qquad\text{(conversion probability)},
\label{eq:ddef}\\[3pt]
\kappa_\ell &\equiv
   \frac{1}{2}\,\frac{\Gax-\Gpol}{\zzeta\,\Gpol}
   \qquad\text{(absorption asymmetry)}.
\label{eq:kappadef}
\end{align}
Both are frequency independent at leading order in $M\omega$; the
remainder of this section reduces them to convergent integrals over
static wavefunctions, and Sec.~\ref{sec:results} evaluates the conversion integral analytically and reconstructs the absorption coefficient to high precision for $\ell=2$.

\subsection{Static (near-zone) solutions}
\label{sec:static}

In the low-frequency regime $M\omega\ll1$ the barrier region is
governed by the $\omega\to0$ (static) limit of
Eq.~\eqref{eq:coupled}. We set $M=1$ throughout this and the
following section; $M$ is restored trivially by dimensional analysis,
and the coefficients $d_\ell,\kappa_\ell$ are dimensionless. For the
quadrupole $\ell=2$ the four static solutions are elementary.
We denote by $\rho$ and $\sigma$ the two linearly independent solutions
of the static ($\omega\to0$) axial Regge--Wheeler equation---the
$\zzeta=0$ limit of the $\Psi$ channel of Eq.~\eqref{eq:coupled}---given
in closed form by the regular solution and its irregular partner,
\begin{align}
\rho(r) &= r^3 ,
\label{eq:rho}\\
\sigma(r) &= -\,r^3\!\int_r^\infty\!\frac{dr'}{f\,r'^{6}}
 = \frac{3r^4\ln\!\frac{r-2}{r}+6r^3+6r^2+8r+12}{96\,r},
\label{eq:sigma}
\end{align}
with $\rho$ regular at the horizon and growing as $r^3$ at infinity,
while $\sigma\sim-1/(5r^2)$ at infinity and is logarithmically
singular at the horizon. Their tortoise-coordinate Wronskian is
constant,
\begin{equation}
W_{\rho\sigma}\equiv
f\big(\rho\,\sigma'-\sigma\,\rho'\big)=+1 .
\label{eq:Wrhosigma}
\end{equation}
Likewise, the two independent solutions $p$ and $q$ of the static
scalar ($\Theta$) channel---the $\zzeta=0$ limit of the second line of
Eq.~\eqref{eq:coupled}---are the associated Legendre functions of
$x\equiv r-1$,
\begin{equation}
p(r)=r\,P_2(r-1),\qquad
q(r)=r\,Q_2(r-1),
\label{eq:pq}
\end{equation}
with $P_2$ regular ($\sim r^3$) and $Q_2$ decaying ($\sim1/r^2$) at
infinity, and Wronskian
\begin{equation}
W_{pq}\equiv f\big(p\,q'-q\,p'\big)=-1 .
\label{eq:Wpq}
\end{equation}
The sign follows from the standard Legendre Wronskian
$\mathcal{W}\{P_2,Q_2\}=-1/(x^2-1)$ with $x=r-1$.

All four solutions are verified to annihilate their static operators
symbolically. In the standard three-zone matching, the horizon and
far zones connect these near-zone forms to purely ingoing and to
spherical-Bessel waves respectively; the general-relativistic
greybody factors follow, with the exact ratio (derived in Appendix~\ref{app:ratio})
\begin{equation}
\frac{\Gamma_{0,\ell}}{\Gamma_{2,\ell}}
=\left[\frac{(\ell-2)!\,(\ell+2)!}{(\ell!)^2}\right]^{-2}
\;\xrightarrow{\ \ell=2\ }\;\frac{1}{36}.
\label{eq:greyratio}
\end{equation}
Here $\Gamma_{0,\ell}$ and $\Gamma_{2,\ell}$ are the spin-0 (scalar) and
spin-2 (axial/polar) general-relativistic greybody factors; at $\ell=2$
we write $\Gamma_s\equiv\Gamma_{0,2}$ and $\Gpol\equiv\Gamma_{2,2}$.

\subsection{Distorted-wave perturbation theory}
\label{sec:dwba}

Because $\Voff\propto\sqrt{\zzeta}$ is concentrated in the near zone,
the conversion and absorption observables are computed by
distorted-wave perturbation theory built on the static solutions
above. At first order in $\Voff$, the pseudoscalar sourced by the incident
graviton is obtained by the standard distorted-wave
(variation-of-parameters) reduction of the coupled
system~\eqref{eq:coupled}: the transmitted conversion amplitude is the
overlap of the incident-graviton solution with the outgoing scalar
mode through the off-diagonal vertex,
\begin{equation}
T_{21}=\frac{T_s}{2i\omega}
       \int \psi_\infty\,\Voff\,\psi_g \; d\rs ,
\label{eq:T21}
\end{equation}
where $\psi_g$ is the incident-graviton distorted wave (near zone
$\propto\rho$), $\psi_\infty$ the scalar up-mode (near zone
$\propto q$), and $T_s$ the scalar greybody amplitude.  The near-zone matching that fixes the normalizations is given in
Appendix~\ref{app:matching}, and the explicit $\ell=2$ reduction of
Eq.~\eqref{eq:T21} to the single integral $J$ is carried out in
Appendix~\ref{app:T21reduction}.

The graviton greybody factor is shifted at second order by the
self-energy generated by integrating out the scalar channel,
$\Sigma=\Voff\,G_s\,\Voff$, where
\begin{equation}
G_s(r,r')=\frac{p(r_<)\,q(r_>)}{W_{pq}}
\label{eq:Gscalar}
\end{equation}
is the static scalar Green's function. Writing the axial transmission
amplitude as $T_{11}=T_g+\delta T_g$, the Wronskian
(inverse-transmission) perturbation gives
\begin{equation}
\frac{\delta T_g}{T_g}
=-\frac{1}{W_1}\!\iint f_L(r)\,\Voff\,G_s(r,r')\,\Voff\,f_R(r')\,
  d\rs\,d\rs' ,
\label{eq:dTg}
\end{equation}
with $f_L$ and $f_R$ the graviton Jost solutions incident from the
horizon and from infinity, and $W_1$ their Wronskian.

The structure of Eq.~\eqref{eq:dTg} organizes the entire
parity-violating response. In the near zone the horizon Jost solution
reduces to the \emph{regular} static solution,
$f_L\to h_g\,\rho$, whereas the infinity Jost solution is dominated by
the \emph{irregular} one, $f_R\to(15/\omega^2)\,\sigma$. The two
legs are thus \emph{different}, and this asymmetry cleanly separates
the physics:
\begin{itemize}
\item the $\rho\times\sigma$ combination is \emph{real}, and gives
the greybody (absorption) asymmetry $\kappa_2$
[Sec.~\ref{sec:kappa}];
\item the $\rho\times\rho$ combination is \emph{imaginary}, and gives
the transmission-\emph{phase} asymmetry [Sec.~\ref{sec:phase}].
\end{itemize}
Using $T_g=2i\omega^3/(15\,h_g)$ and $W_1=15\,h_g/\omega^2$ (derived in Appendix~\ref{app:matching}), the
$\omega$- and normalization-dependent factors in
Eq.~\eqref{eq:dTg} cancel identically, leaving frequency-independent
coefficients determined purely by the static integrals. In brief: working to $\Ord(\sqrt{\zzeta})$ on the static $\ell=2$ basis
$\{\rho,\sigma;p,q\}$, the first-order conversion $T_{21}$
[Eq.~\eqref{eq:T21}] follows from one off-diagonal vertex and the axial
transmission correction [Eq.~\eqref{eq:dTg}] from two vertices closed by
the scalar Green's function; the far-zone normalizations are fixed by
matched asymptotics (Appendix~\ref{app:matching}) and the results
validated against direct coupled-channel integration
(Sec.~\ref{sec:checks}).

\section{Analytic Conversion and High-Precision Absorption Coefficients}
\label{sec:results}

We now evaluate the static integrals of Sec.~\ref{sec:dwba} for the
quadrupole $\ell=2$. The conversion coefficient is obtained analytically in closed form, while the absorption coefficient is reconstructed to high precision within the analytically controlled basis described below. Throughout, $x\equiv r-1$ and
$P_2(x)=(3x^2-1)/2$,
$Q_2(x)=\tfrac12 P_2(x)\ln\frac{x+1}{x-1}-\tfrac32 x$.

\subsection{Conversion probability \texorpdfstring{$d_2$}{d2}}
\label{sec:d2}

Inserting the near-zone forms $\psi_g\to\rho=r^3$ and
$\psi_\infty\to q=rQ_2$ into Eq.~\eqref{eq:T21} and using
$\Voff=12\sqrt6\,\sqrt{\zzeta}\,f/r^5$ (i.e.\ $6\sqrt\Lambda$ with
$\Lambda=24$), the conversion amplitude collapses to a single
integral,
\begin{equation}
T_{21}=-18\sqrt6\,\sqrt{\zzeta}\;T_s\,J,
\qquad
J=\int_1^\infty\frac{Q_2(x)}{x+1}\,dx .
\label{eq:T21J}
\end{equation}
The integrand reduces to elementary logarithms plus a single
dilogarithm; evaluating the endpoints (Appendix~\ref{app:Jvalue}) gives
$+\pi^2/12$ from the dilogarithm and $-3/4$ from the
polynomial--logarithmic part, hence
\begin{equation}
J=\frac{\pi^2}{12}-\frac34=\frac{\pi^2-9}{12}
\label{eq:J}
\end{equation}
(exact to $30$ digits against numerical quadrature). Squaring the
amplitude and normalizing by the polar greybody factor with the exact
ratio $\Gamma_s/\Gpol=1/36$ from Eq.~\eqref{eq:greyratio},
\begin{equation}
d_2=\frac{\Ggs}{\zzeta\,\Gpol}
   =(18\sqrt6)^2\,\frac{1}{36}\,J^2
   =54\,J^2 ,
\end{equation}
so that
\begin{equation}
d_2=\frac{3(\pi^2-9)^2}{8}=0.283579430\ldots\;
\label{eq:d2}
\end{equation}
Every factor entering Eq.~\eqref{eq:d2} is exact: $J$ from the
symbolic antiderivative, the vertex factor $(18\sqrt6)^2=1944$, and
the greybody ratio $1/36$. The value agrees with a full two-channel
numerical integration of Eq.~\eqref{eq:coupled} through the
frequency-independent ratio tests of Appendix~\ref{app:d2}.

\subsection{Absorption asymmetry \texorpdfstring{$\kappa_2$}{kappa2}}
\label{sec:kappa}

For the greybody asymmetry, the $\rho\times\sigma$ leg structure of
Eq.~\eqref{eq:dTg} reduces the double integral to
\begin{equation}
\begin{aligned}
    \kappa_2 &= -\,864\,D_2^{\kappa},\\
D_2^{\kappa} &=\int_2^\infty
   \frac{\sigma(r)}{r^{5}}\big[\,q(r)\,A(r)+p(r)\,B(r)\,\big]\,dr ,
\end{aligned}
\label{eq:kappaint}
\end{equation}
with the prefactor $864=(12\sqrt6)^2$ from the two conversion vertices
and the Wronskians of Eqs.~\eqref{eq:Wrhosigma},\eqref{eq:Wpq}. The
inner integrals are the elementary
\begin{equation}
A(r)=\int_2^{r}\frac{P_2(r'-1)}{r'}\,dr'
    =\frac{3r^2}{4}-3r+\ln r+3-\ln 2 ,
\label{eq:A}
\end{equation}
and $B(r)=H(\infty)-H(r-1)$, where $H(x)=\int\!\frac{Q_2(x)}{x+1}dx$
is the same dilogarithmic antiderivative as in
Eq.~\eqref{eq:T21J}, with the exact limit
\begin{equation}
H(\infty)=-\frac{21}{16}
\label{eq:Hinf}
\end{equation}
(the explicit $H$ is given in Appendix~\ref{app:kappa}).

The integrand of Eq.~\eqref{eq:kappaint} has transcendental weight
three, so $D_2^{\kappa}$ is a weight-four constant; the weight argument
(Appendix~\ref{app:weight}) fixes the candidate basis, and an
integer-relation (PSLQ) search verified to $56$ digits
(Appendix~\ref{app:pslq}) identifies it as an even-weight combination
of $\pi^4$, $\pi^2$ and a rational number, the odd-zeta and logarithmic
constants cancelling:
\begin{equation}
\kappa_2=\frac{21}{80}\pi^4+\frac{27}{8}\pi^2-\frac{89833}{1536}
       =0.394775208\ldots
\label{eq:kappa2}
\end{equation}
equivalently $(2016\,\pi^4+25920\,\pi^2-449165)/7680$. This matches
the direct two-channel numerical greybody asymmetry at the accessible low-frequency precision.

\subsection{Unitarity check and transcendental structure}
\label{sec:structure}

The result~\eqref{eq:kappa2} admits a transparent check through flux
conservation: writing $\kappa_2=-\tfrac12 d_2+\kappa_R$, the
conversion-loss term $-\tfrac12 d_2=-27J^2$ supplies the \emph{entire}
$\pi^2$ content of $\kappa_2$, leaving the dispersive reflection term
$\kappa_R=\kappa_2+27J^2=\tfrac{9}{20}\pi^4-\tfrac{66505}{1536}$ as a
pure quartic with no $\pi^2$---an independent consistency check
(Appendix~\ref{app:unitarity}).

The two absorption-sector coefficients thus reduce to combinations involving only even powers of $\pi$ and rational terms, while the \emph{dispersive} phase sector
(Sec.~\ref{sec:phase}) carries the odd Ap\'ery constant $\zeta(3)$.
This even/odd split follows the reality structure of the
$\rho\times\sigma$ (real, absorptive) versus $\rho\times\rho$
(imaginary, dispersive) leg combinations identified below
Eq.~\eqref{eq:dTg}. Table~\ref{tab:coeffs} collects the coefficients.
\squeezetable
\begin{table}[t]
\caption{Leading $\Ord(\zzeta)$ parity-violating coefficients for the
$\ell=2$ channel of a Schwarzschild \dCS{} black hole. The absorption sector reduces to a $\pi^4$, $\pi^2$, and rational combination; the dispersive static phase integral contains $\zeta(3)$.}
\label{tab:coeffs}
\begin{ruledtabular}
\begin{tabular}{lccc}
Observable & Symbol & Result & Value \\
\hline
Conversion            & $d_2$      & $\dfrac{3(\pi^2-9)^2}{8}$
   & $0.283579$\\[6pt]
Absorption asymmetry  & $\kappa_2$ &
   $\dfrac{21\pi^4}{80}+\dfrac{27\pi^2}{8}-\dfrac{89833}{1536}$
   & $0.394775$\\[6pt]
Transmission phase    & $c_1$ &
   $2\pi^2\big(\zeta(3)-\tfrac98\big)$\footnote{The factor
   $\zeta(3)-9/8$ is derived analytically (Appendix~\ref{app:phase});
   the prefactor $2\pi^2$ is a numerical identification, to four digits.}
   & $1.52104$\\
Reflection phase (PV) & $\Delta\phi_R$ & $\Ord\!\big(\zzeta(M\omega)^5\big)$ & --- \\
\end{tabular}
\end{ruledtabular}
\end{table}

\begin{figure*}[t]
\centering
\includegraphics[width=0.98\textwidth]{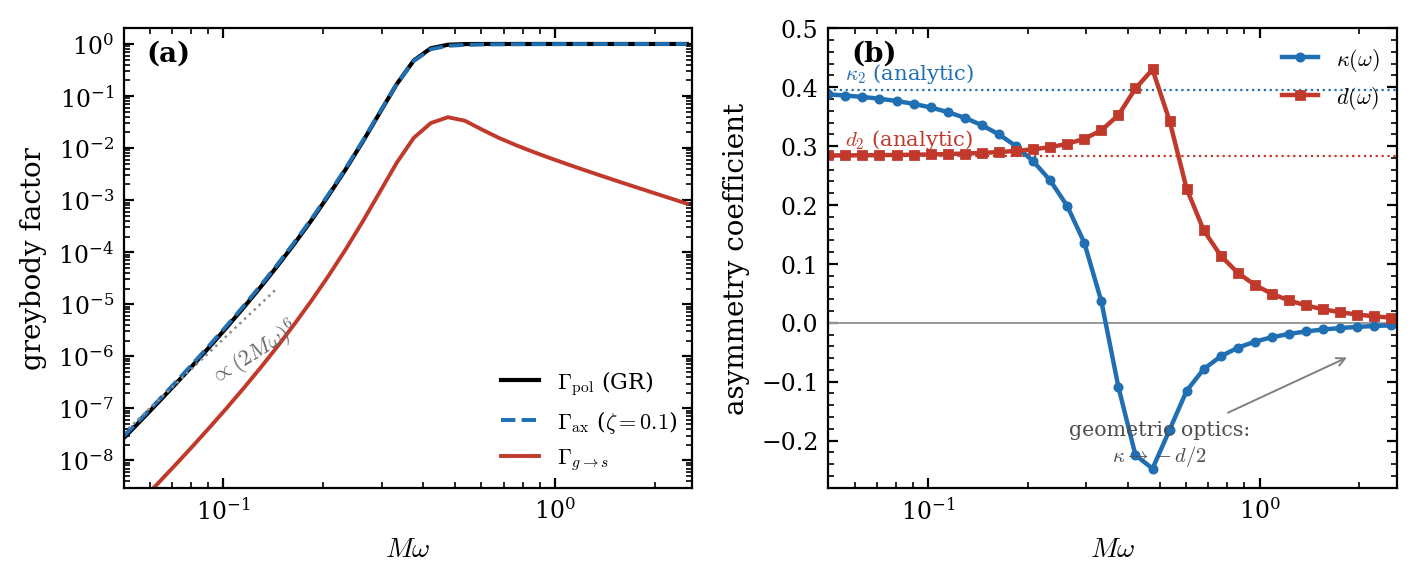}
\caption{(a) Greybody factors of the $\ell=2$ channel versus $M\omega$
(coupling $\zzeta=0.1$) from numerical integration of the coupled
system~\eqref{eq:coupled}: the polar (general-relativistic) $\Gpol$,
the axial $\Gax$, and the graviton$\to$pseudoscalar conversion
$\Ggs$. All obey the low-frequency law $\propto(2M\omega)^6$ (dotted
guide) and approach unity at high frequency; $\Gax$ and $\Gpol$ are
nearly degenerate, their difference being $\Ord(\zzeta)$. (b) The
parity-violating asymmetry coefficients
$\kappa(\omega)=\tfrac12(\Gax-\Gpol)/(\zzeta\Gpol)$ and
$d(\omega)=\Ggs/(\zzeta\Gpol)$, extrapolated to $\zzeta\to0$. As
$M\omega\to0$ they approach the analytic forms $\kappa_2$
[Eq.~\eqref{eq:kappa2}] and $d_2$ [Eq.~\eqref{eq:d2}] (dotted lines),
the small residual at the lowest sampled frequencies being the
$\Ord((M\omega)^2)$ correction. At high frequency the asymmetry is
purely absorptive and geometric-optics suppressed,
$\kappa(\omega)\to-d(\omega)/2$ [i.e.\ $\Gax-\Gpol=-\Ggs$], with both
$\to0$ as $(M\omega)^{-2}$.}
\label{fig:greybody}
\end{figure*}

\section{Physical observables and signatures}
\label{sec:phys}

The coefficients $d_2$ and $\kappa_2$ control three physically
distinct imprints of parity violation on wave propagation past the
black hole: a parity-dependent absorption, an emission of
pseudoscalar radiation, and---through the phase sector---a strongly suppressed reflection-phase birefringence. We
discuss each in turn, and close with the vanishing of the net
thermal circular polarization.

\subsection{Absorption asymmetry and induced ellipticity}
\label{sec:absorption}

The sign $\kappa_2>0$ in Eq.~\eqref{eq:kappa2} means that the axial
graviton is absorbed \emph{more strongly} than its polar counterpart,
$\Gax>\Gpol$ at $\Ord(\zzeta)$. The mechanism is transparent in the
adiabatic (Born--Oppenheimer) picture of the coupled system. Since
the diagonal gap $V_{11}-V_{22}=-8Mf/r^3$ never vanishes, the
$2\times2$ potential matrix of Eq.~\eqref{eq:coupled} may be
diagonalized locally, and the graviton eigen-potential is depressed
by level repulsion,
\begin{equation}
\delta V_1=-\frac{\Voff^2}{V_{22}-V_{11}}
          =-108\,\zzeta\,\frac{f}{r^{7}} \; (M=1),
\label{eq:deltaV}
\end{equation}
a \emph{lowered} barrier and hence enhanced transmission. (The full
$\kappa_2$ requires the nonlocal Green's function of
Sec.~\ref{sec:dwba}; the adiabatic estimate captures the sign and
the strong-field localization but not the coefficient.)

Because the two linear polarizations of an incident gravitational
wave map onto the axial and polar master variables, a parity-even
absorption difference $\Gax\neq\Gpol$ acts as a weak, frequency- and
geometry-dependent \emph{dichroism}: the transmitted/absorbed field
acquires a small induced ellipticity of order
$\kappa_2\,\zzeta\,\Gpol$ relative to the general-relativistic case.
This is the strong-field analogue of the birefringent absorption
familiar from parity-violating optics, now sourced by the
gravitational Pontryagin coupling rather than by a medium.

Concretely, in a coherent scattering geometry the two linear master
amplitudes propagate diagonally,
\begin{equation}
\begin{pmatrix}A_{\rm pol}^{\rm out}\\ A_{\rm ax}^{\rm out}\end{pmatrix}
=\begin{pmatrix}T_{\rm pol}&0\\ 0&T_{\rm ax}\end{pmatrix}
 \begin{pmatrix}A_{\rm pol}^{\rm in}\\ A_{\rm ax}^{\rm in}\end{pmatrix},
 \label{eq:prop_matrix}
\end{equation}
so unequal magnitudes $|T_{\rm ax}|\neq|T_{\rm pol}|$ produce
differential attenuation (the induced ellipticity above), while a
relative phase would rotate or reshape the polarization ellipse of a
coherent wave. The reflected relative phase is strongly suppressed at low frequency [Eq.~\eqref{eq:norotation}], so negligible rotation is generated there; the magnitude of the induced ellipticity
depends on the incident state and the observation geometry.

\subsection{Pseudoscalar emission}
\label{sec:emission}

The conversion probability $d_2$ quantifies an irreducibly
gravitational process: an incident axial gravitational wave is partly converted into propagating \dCS{} pseudoscalar radiation, which is then absorbed by
(or, in the time-reverse, emitted from) the black hole,
\begin{equation}
\Ggs=d_2\,\zzeta\,\Gpol
    =\frac{3(\pi^2-9)^2}{8}\,\zzeta\,\Gpol .
\end{equation}
By detailed balance, the same coupled-channel transmission factors enter the Hawking-emission problem: the pseudoscalar channel contributes to the emitted spectrum whenever the corresponding massless scalar mode is kinematically open, in fixed ratio $d_2\,\zzeta$ relative to the axial-graviton channel.
For the massless field considered here the channel is open at all
frequencies; a pseudoscalar mass $m_\vartheta$ (the axion-like /
dark-matter extension) introduces a threshold $\omega>m_\vartheta$
below which conversion is kinematically forbidden, a point we return
to in Sec.~\ref{sec:discussion}.

\subsection{Phase sector: reflection and transmission phases}
\label{sec:phase}

In general relativity the axial (Regge--Wheeler) and polar (Zerilli)
channels are isospectral: they share identical transmission and reflection
moduli, $|T^{\rm RW}_\ell|=|T^{\rm Z}_\ell|$ and
$|R^{\rm RW}_\ell|=|R^{\rm Z}_\ell|$.  Their reflection \emph{phases},
however, differ by a nonzero, parity-even amount fixed entirely by the
intertwining (Chandrasekhar--Detweiler) relation between the two
potentials~\cite{Chandrasekhar1983,Glampedakis2017},
\begin{equation}
\begin{aligned}
    \arg R^{\rm RW}_\ell-\arg R^{\rm Z}_\ell
  &=-2\arctan\!\left[\frac{12\,M\omega}
       {(\ell-1)\ell(\ell+1)(\ell+2)}\right]\\
  &\;\xrightarrow[\ \ell=2\ ]{}\;
  -2\arctan\!\frac{M\omega}{2}\;\simeq\;-\,M\omega .
\end{aligned}
\label{eq:CD}
\end{equation}
This offset is $\Ord(M\omega)$, is independent of $\zzeta$, and is already
present in GR; it is \emph{not} a signature of parity violation.  We
confirmed Eq.~\eqref{eq:CD} numerically to all available digits by
integrating the Regge--Wheeler and Zerilli problems directly (e.g.\ at
$M\omega=0.2$, $-2\arctan0.1=-0.199330$ against a measured $-0.199295$).

Because the polar sector reduces to GR at every order in $\zzeta$, the
only parity-violating effect in the reflected wave is the $\Ord(\zzeta)$
shift of the \emph{axial} reflection phase about its GR value,
\begin{equation}
  \Delta\phi_R\;\equiv\;
  \arg R^{\rm ax}_{11}-\arg R^{\rm RW}_{11}
  \;=\;\Ord\!\big(\zzeta\,(M\omega)^5\big),
  \label{eq:norotation}
\end{equation}
numerically $\Delta\phi_R\propto\zzeta\,(M\omega)^5$ (up to a possible
logarithm) across the accessible range $M\omega\lesssim0.2$, and vanishing
as $M\omega\to0$.  (In the submitted version this quantity was obtained
with the Regge--Wheeler potential as a polar proxy---exact for the
greybody \emph{moduli}, and hence for $d_2$ and $\kappa_2$, but insensitive
to the $\Ord(M\omega)$ GR phase of Eq.~\eqref{eq:CD}; the parity-violating
shift is the baseline-subtracted quantity~\eqref{eq:norotation}.)

The phase sector is thus ordered by three disparate scales: the
parity-\emph{even} GR reflection phase of Eq.~\eqref{eq:CD},
$\Ord(M\omega)$; the parity-\emph{odd} reflection-phase birefringence
$\Delta\phi_R=\Ord(\zzeta\,(M\omega)^5)$; and the parity-odd
\emph{transmission} phase, whose coefficient
$c_1=2\pi^2\big(\zeta(3)-\tfrac98\big)=\Ord(\zzeta\,\omega)$
[Table~\ref{tab:coeffs}] is the leading dispersive imprint of \dCS{}
parity violation.  The reflected-wave birefringence is suppressed relative
to the transmission channel by four powers of $M\omega$ and is negligible
throughout the low-frequency regime in which our closed forms hold.
 
We stress that the axial and polar master variables are gauge-invariant
combinations of the metric perturbation and are \emph{not} themselves the
$+$ and $\times$ strain polarizations, which are specific angular
superpositions of the two~\cite{MartelPoisson2005}.  The leading
polarization-space signature of the unequal parity-channel greybodies is
accordingly a differential attenuation---an induced ellipticity from
$|T_{\rm ax}|\neq|T_{\rm pol}|$---rather than a rotation of the
polarization ellipse; the latter is controlled by the parity-odd phase
difference $\Delta\phi_R$, which Eq.~\eqref{eq:norotation} shows to be
strongly suppressed at low frequency.

The \emph{transmission}-phase asymmetry---the relative phase
of the into-horizon amplitudes---is nonzero but subleading, growing
linearly with frequency,
\begin{equation}
\arg T_{11}^{\rm ax}-\arg T_{11}^{\rm pol}
   \simeq c_1\,\omega\,\zzeta,\qquad
c_1\simeq 2\pi^2\Big(\zeta(3)-\tfrac98\Big),
\label{eq:transphase}
\end{equation}
where the transcendental factor $\zeta(3)-9/8$ is itself
\emph{derived in closed form}: it is the $\rho\times\rho$ partner of
Eq.~\eqref{eq:kappaint}, a weight-three non-alternating polylogarithm
that evaluates to $\zeta(3)-9/8$ by the same static-integral reduction
used for $\kappa_2$ (Appendix~\ref{app:phase}). Only the overall
prefactor $2\pi^2$ remains a numerical identification:
$c_1/[\zeta(3)-9/8]\to 2\pi^2$ to four digits in the two-channel
integration, while a first-principles derivation of $2\pi^2$---which
requires the dispersive near/far-zone matching---is left to future work. We stress that Eq.~\eqref{eq:transphase} affects only
the absorbed radiation and is subdominant to the observables of
Secs.~\ref{sec:absorption}--\ref{sec:emission}; the physically salient phase statement is the strong low-frequency
suppression of the reflection-phase birefringence~\eqref{eq:norotation}.

\subsection{Absence of net thermal circular polarization}
\label{sec:hawking}

It is tempting to interpret $\Gax\neq\Gpol$ as a net circular
polarization of the black hole's own (Hawking) emission. It is not.
The circular-polarization basis is the parity combination
$|R/L\rangle=(|{\rm pol}\rangle\pm i\,|{\rm ax}\rangle)/\sqrt2$, so a
greybody difference between the two parity eigenmodes does not by
itself bias the thermal populations: the parity eigenstates are
populated independently and equally, and the cross term
$\langle\psi_{\rm pol}\,\psi_{\rm ax}^{*}\rangle$ vanishes upon
angular integration of a spherically symmetric background. Hence, for an unpolarized, spherically symmetric thermal state with no parity-coherent cross correlations, the net Hawking circular polarization vanishes for the non-rotating \dCS{} black hole, despite the unequal parity-channel greybody factors. A nonzero effect requires a preferred axis, i.e.\ black
hole spin; the Kerr generalization is the natural setting for a net
chiral emission and is discussed in Sec.~\ref{sec:discussion}.

\section{Limits, checks, and multipole dependence}
\label{sec:checks}

\subsection{Consistency checks}
\label{sec:consistency}

The analytic coefficients pass a series of independent checks. In the
limit $\zzeta\to0$ the axial channel reduces to Regge--Wheeler and its
greybody factor coincides with the polar Zerilli value, restoring
isospectrality: $\Gax\to\Gpol$, $\Ggs\to0$, and hence
$\kappa_2,d_2$ enter strictly at $\Ord(\zzeta)$. The polar sector is
general-relativistic to all orders in $\zzeta$, as it must be, since
$\vartheta$ does not source even-parity perturbations. Flux
conservation
$|R_{11}|^2+|R_{21}|^2+|T_{11}|^2+|T_{21}|^2=1$ is satisfied
throughout the two-channel integration and underlies the unitarity
decomposition of Sec.~\ref{sec:structure}.

Numerically, we integrate Eq.~\eqref{eq:coupled} with purely ingoing
horizon data and extract $\Gax$, $\Ggs$, and $\Gpol$; the results are
shown in Fig.~\ref{fig:greybody}. Panel~(a) displays the three
greybody factors, all obeying the low-frequency law
$\propto(2M\omega)^6$, with $\Gax$ and $\Gpol$ nearly degenerate.
Panel~(b) isolates their $\Ord(\zzeta)$ difference through the
coefficients $\kappa(\omega)$ and $d(\omega)$
[Eqs.~\eqref{eq:kappadef},\eqref{eq:ddef}] extrapolated to
$\zzeta\to0$: the frequency-independent, $\zzeta$-linear combinations
approach $d_2=0.283579$ and $\kappa_2=0.394775$ as $M\omega\to0$, in
agreement with Eqs.~\eqref{eq:d2} and \eqref{eq:kappa2}. The analytic $d_2$ result and the PSLQ reconstruction of $\kappa_2$ are numerically verified to 30 and 56 digits, respectively, by
the symbolic and integer-relation analyses of
Appendices~\ref{app:d2}--\ref{app:kappa}. 

The finite-$\zzeta$ curves shown here are solutions of the
perturbatively truncated master system of Eq.~\eqref{eq:coupled}---
retaining the $\Ord(\sqrt{\zzeta})$ off-diagonal coupling and the
$\Ord(\zzeta)$ diagonal correction---rather than of the full
nonperturbative \dCS{} theory. The physical $\Ord(\zzeta)$ coefficients
$d_\ell,\kappa_\ell$ are obtained from the small-$\zzeta$, low-frequency
behavior extrapolated to $\zzeta\to0$ and $M\omega\to0$; the
representative value $\zzeta=0.1$ used in the figures is chosen only to
render the $\Ord(\zzeta)$ effects visible.

For completeness, Appendix~\ref{app:smatrix} assembles the full
two-channel S-matrix at $\ell=2$, including the pseudoscalar-incidence
column---the massless-scalar greybody and the scalar$\to$graviton
conversion. It satisfies unitarity, $R^\dagger R+T^\dagger T=\mathbb{1}$,
and reciprocity, $S=S^{\mathsf T}$, to $\sim10^{-12}$. The
parity-violating coefficients $d_2$ and $\kappa_2$ reside entirely in the
graviton-incidence column and are unaffected by the scalar-incidence
sector; the same block structure and $(2M\omega)^{2\ell+2}$ threshold
scalings hold for all $\ell\ge2$, with the effect dominated by $\ell=2$.

\begin{table*}[htbp]
\caption{Status of the principal results: what is established by exact
analysis, what is an exact reduction evaluated to high precision, what is
reconstructed by an integer-relation (PSLQ) search, and what is a direct
numerical result or fit. The physics is unchanged; this makes the
evidentiary hierarchy explicit.}
\label{tab:status}
\begin{ruledtabular}
\begin{tabular}{p{0.42\textwidth}p{0.52\textwidth}}
Quantity & Status \\ \hline
Static solutions $\rho,\sigma,p,q$ [Eqs.~\eqref{eq:rho}--\eqref{eq:pq}] & Exact analytic \\
Wronskians $W_{\rho\sigma},\,W_{pq}$ & Exact analytic \\
Matching factors $h_g,W_1,T_g,T_s$ and ratio $\Gamma_s/\Gpol=1/36$ & Exact analytic \\
Conversion integral $J=(\pi^2-9)/12$ & Exact analytic \\
Conversion coefficient $d_2$ [Eq.~\eqref{eq:d2}] & Exact analytic closed form \\
Self-energy reduction $\to D_2^{\kappa}$ & Exact analytic reduction to a one-dimensional integral \\
$D_2^{\kappa}$ [Eq.~\eqref{eq:D2knested}] & One-dimensional integral; high-precision quadrature ($58$ digits) \\
Absorption asymmetry $\kappa_2$ [Eq.~\eqref{eq:kappa2}] & High-precision PSLQ reconstruction in analytically controlled basis; exact form strongly supported ($\zeta(3)$ coefficient $=0$) \\
Phase integral $D_2=\zeta(3)-9/8$ [Eq.~\eqref{eq:D2zeta}] & Exact analytic \\
Transmission-phase coefficient $c_1$ [Eq.~\eqref{eq:transphase}] & Analytic static factor $\times$ numerically identified prefactor $2\pi^2$ \\
Reflection-phase asymmetry [Eq.~\eqref{eq:norotation}] & Numerical low-frequency result (null at leading order) \\
High-frequency coefficient $c_\infty$ & Numerical fit \\
$\ell=3$ coefficients $d_3,\kappa_3$ & Numerical low-frequency extrapolation \\
\end{tabular}
\end{ruledtabular}
\end{table*}

\subsection{Geometric-optics limit}
\label{sec:geo}

The coefficients $d_\ell,\kappa_\ell$ are defined in the
low-frequency regime; the opposite, geometric-optics limit
$M\omega\to\infty$ provides a complementary check and clarifies the
character of the effect. There all greybody factors approach unity,
and the parity-violating asymmetry becomes purely absorptive: the
dispersive barrier shift switches off and the graviton greybody
deficit is exactly the flux lost to conversion,
\begin{equation}
\Gax-\Gpol \;=\; -\,\Ggs
\qquad (M\omega\to\infty),
\label{eq:geo}
\end{equation}
with both sides falling as
\begin{equation}
\frac{\Gax-\Gpol}{\zzeta\,\Gpol}
=-\frac{\Ggs}{\zzeta\,\Gpol}
\;\sim\; -\,\frac{c_\infty}{(M\omega)^2},
\quad c_\infty\simeq0.053 .
\end{equation}
The high-frequency coefficient is a numerical fit, $c_\infty=0.0528(2)$ for $\ell=2$ over $M\omega\in[2,10]$, and the identity~\eqref{eq:geo} holds to five digits.
This is visible in Fig.~\ref{fig:greybody}(b): beyond the
low-frequency plateau the asymmetry coefficient $\kappa(\omega)$
changes sign and tends to $-d(\omega)/2$, the signature of
Eq.~\eqref{eq:geo}. The numerically observed $1/(M\omega)^2$ falloff indicates that the parity-violating imprint is an intrinsically strong-field, low-frequency phenomenon: at high
frequency the wave sails over the barrier and the only surviving
$\Ord(\zzeta)$ effect is the (small) conversion loss. This places the
observables of this paper squarely in the regime probed by the
ringdown and by low-frequency scattering, rather than by the
eikonal/photon-sphere sector. The low-frequency strong-field scattering regime studied here is
complementary to, and may numerically overlap with, the dimensionless
frequency range $M\omega=\Ord(1)$ relevant to black-hole ringdown,
rather than the eikonal/photon-sphere sector; raising the black-hole mass
lowers the absolute frequency at fixed $M\omega$ but does not make
$M\omega$ parametrically smaller.

\subsection{Multipole dependence}
\label{sec:multipole}

The quadrupole $\ell=2$ dominates gravitational-wave phenomenology,
and it is the case for which we have obtained analytic results. The
construction extends to higher multipoles with no change of
principle: the general-relativistic greybody ratio entering
Eq.~\eqref{eq:greyratio} is
\begin{equation}
\begin{aligned}
\frac{\Gamma_{0,\ell}}{\Gamma_{2,\ell}}
&= \left[\frac{(\ell-2)!\,(\ell+2)!}{(\ell!)^2}\right]^{-2} \\[4pt]
&= \frac{1}{36},\;\frac{9}{100},\;\ldots \quad (\ell=2,3,\ldots).
\end{aligned}
\end{equation}
and the coupling strength grows through $\Lambda=(\ell+2)!/(\ell-2)!$
in Eq.~\eqref{eq:Voff}. Evaluating the two-channel system numerically
for $\ell=3$ we obtain the larger normalized coefficients,
\begin{equation}
d_3=0.87(1),\qquad \kappa_3=0.85(1) ,
\end{equation}
(ratio method, $M\omega\in[0.03,0.06]$, $\zzeta=0.01$ and $0.02$; the quoted uncertainty is a conservative aggregate dominated by the fit-window and $\zzeta$-extrapolation variation), to be compared with $d_2=0.2836$, $\kappa_2=0.3948$. This single additional multipole suggests the effect becomes larger, in normalized units, for higher
multipoles, though it does not by itself establish a general trend; any such growth is offset in a physical observable by the smaller
contribution of higher multipoles to the radiated signal; closed forms for general $\ell$, and the associated spectral sums, are a natural extension left to future work.

\section{Discussion and outlook}
\label{sec:discussion}

\subsection{Relation to the quasinormal-mode spectrum}
\label{sec:qnm}

The greybody asymmetry $\kappa_2$ and the axial/polar quasinormal-mode
(QNM) splitting of \dCS{} gravity are two facets of the same
isospectrality breaking. In general relativity the Regge--Wheeler and
Zerilli potentials are isospectral, so axial and polar perturbations
share both their QNM frequencies and their greybody factors; the
\dCS{} coupling lifts this degeneracy at $\Ord(\zzeta)$. The QNM
splitting has been computed by
Refs.~\cite{CardosoGualtieri2009,Molina2010,WagleYunesSilva2021};
the present work supplies the complementary transmission-sector
statement, in closed form, for the same coupled system. Because the
greybody factors and QNMs are controlled by the same barrier~\cite{KonoplyaZhidenko2024,PedrottiCalza2025,Antoniou2026}, the absorption asymmetry~\eqref{eq:kappa2} and the ringdown frequency
shift are complementary manifestations of the same dCS-induced lifting of the axial/polar degeneracy, arising from the same coupled system and barrier physics; the present work does not, however, derive a direct quantitative map between the two observables. The scattering observables are, moreover, sensitive
to a regime the QNMs are not---the graviton$\to$pseudoscalar
conversion channel~\eqref{eq:d2}, which has no analogue in the
single-channel ringdown of the metric sector alone.

\subsection{Relation to the tidal response}
\label{sec:love}

The low-frequency scattering problem solved here shares its near-zone
content with the tidal-response (Love-number) problem for the same
black hole. The static axial solutions $\rho=r^3$ and $\sigma$
[Eqs.~\eqref{eq:rho}--\eqref{eq:sigma}] are precisely the growing
(tidal) and decaying (response) solutions whose coefficient ratio
defines the gravitational Love number, while the greybody factor is
the leading \emph{dissipative} part of the same low-frequency
response---the black-hole tidal dissipation number. The conservative
tidal Love numbers of non-spinning \dCS{} black holes were computed in
Ref.~\cite{CardosoFranzin2017}; the coefficients $\kappa_2$ and $d_2$
obtained here are the complementary, $\Ord(\zzeta)$ dissipative
response coefficients of the axial sector, built from the identical
static basis $\{\rho,\sigma;p,q\}$. In the language of the worldline
effective field theory of compact objects, the greybody and conversion
coefficients fix the leading parity-violating \emph{absorptive}
couplings---the dissipative counterparts of the conservative Love-number
couplings---placing the present results within the same matching
programme~\cite{CassemHertzberg2026}.

\subsection{Observational regime and magnitude}
\label{sec:regime}

The size of every effect is set by $\zzeta=(\ell_{\rm CS}/M)^4$, and
the effective-field-theory validity of \dCS{} gravity requires
$\ell_{\rm CS}\lesssim M$, i.e.\ $\zzeta\lesssim1$~\cite{CassemHertzberg2026}, while the perturbative expansion used here additionally requires $\zzeta\ll1$. Current bounds on $\ell_{\rm CS}$ from the gravitational-wave and neutron-star sectors
therefore make $\zzeta$ appreciable only for the lightest black
holes, where $GM/c^2$ is comparable to $\ell_{\rm CS}$; for
supermassive holes $\zzeta$ is minuscule. Combined with the
geometric-optics suppression of Sec.~\ref{sec:geo}, which confines
the asymmetry to $M\omega\lesssim1$, the natural observational window
is the strong-field, low-frequency response of stellar-mass and
intermediate-mass black holes---the ringdown and near-threshold
scattering regime---rather than the high-frequency tail. We do not
claim present detectability; rather, the value of the closed forms is
that they convert the qualitative expectation ``axial and polar
respond differently'' into definite $\Ord(\zzeta)$ targets
($d_2,\kappa_2$ of order $0.3$--$0.4$) that any data analysis or
numerical-relativity study can calibrate against without a free
normalization.

\subsection{Extensions}
\label{sec:extensions}

Three extensions follow naturally.

\emph{Massive / dark-sector pseudoscalar.} Restoring a mass
$m_\vartheta$ in Eq.~\eqref{eq:action} gates the conversion channel at
$\omega>m_\vartheta$ and turns the emitted \dCS{} scalars into a
candidate light dark-sector field. The conversion
coefficient~\eqref{eq:d2} then sets the branching into this channel
above threshold, connecting the parity-violating gravitational sector
to dark-matter phenomenology; a quantitative treatment requires the
massive scalar Green's function in place of
Eq.~\eqref{eq:Gscalar}.

\emph{General multipoles and spectral sums.} The $\ell=2$ closed
forms extend in principle to all $\ell$ through the same static
solutions and integer-relation identification
(Sec.~\ref{sec:multipole}); a closed general-$\ell$ family would
permit the spectral sums entering the total absorbed flux and the
scalar luminosity.

\emph{Rotation.} The vanishing of the net thermal circular
polarization (Sec.~\ref{sec:hawking}) and of the scattered-wave
rotation (Sec.~\ref{sec:phase}) are both consequences of spherical
symmetry. Black-hole spin supplies the preferred axis needed for a
genuine chiral imprint, and the frame-dragged, oblate coupling
geometry is expected to convert the parity-even absorption asymmetry
found here into a net helicity effect. The Kerr generalization---where
the strong-field parity violation of \dCS{} gravity would couple to
the spin and to any surrounding matter distribution---is the most
promising direction for turning these analytic results into an
observational diagnostic.

\section{Conclusions}
\label{sec:conclusions}

We have given a first-principles account of the leading parity-violating imprint of dynamical Chern–Simons gravity on low-frequency wave scattering off a Schwarzschild black hole. The quadrupolar conversion coefficient is obtained analytically in closed form, while the absorption asymmetry is reduced to a one-dimensional static integral and reconstructed to high precision within an analytically controlled basis: the
graviton$\to$pseudoscalar conversion probability
$d_2=3(\pi^2-9)^2/8$ and the graviton greybody asymmetry
$\kappa_2=\tfrac{21}{80}\pi^4+\tfrac{27}{8}\pi^2-\tfrac{89833}{1536}$,
both reducible to combinations of $\pi^4$, $\pi^2$, and rational terms, of order a few tenths. A single
structural feature---that the two graviton legs of the second-order
self-energy are the regular ($r^3$) and irregular ($\sigma$) static
solutions---organizes the entire response, separating the real,
absorptive sector (the $\pi^4$ coefficients above) from the
dispersive phase sector, which instead carries Ap\'ery's constant
$\zeta(3)$. The parity-violating reflection-phase asymmetry is strongly suppressed
at low frequency---far below the parity-even general-relativistic
reflection phase---so the reflected gravitational wave suffers
negligible polarization rotation,
and the net thermal (Hawking) circular polarization is likewise zero
for the non-rotating hole; both are consequences of spherical
symmetry. In the geometric-optics regime the numerical results are consistent with a
$(M\omega)^{-2}$ falloff of the normalized flux asymmetries, confirming that the effect is a genuinely
strong-field, low-frequency one, and leaving only the conversion loss
$\Gax-\Gpol=-\Ggs$ at high frequency. The matched-asymptotic coefficients are checked against direct two-channel integration at the accessible low-frequency precision, while
$\kappa_2$ is independently reconstructed to high precision from its
reduced one-dimensional integral within an analytically controlled basis.

These quantitative coefficients turn a qualitative expectation into normalization-fixed predictions, correlated with---but distinct
from---the \dCS{} ringdown. The most compelling directions ahead are
the inclusion of a pseudoscalar mass, which gates the conversion
channel and links it to dark-sector physics; the extension to general
multipoles and the associated spectral sums; and, above all, the
generalization to a rotating (Kerr) black hole, where spin provides
the preferred axis needed to convert the parity-even absorption
asymmetry found here into a net chiral, potentially observable,
signature.

\section*{Data Availability Statement}
No new experimental data were generated in this study. All computational
scripts required to reproduce the figures and numerical results are
provided in the Supplemental Material and archived at Zenodo, version v1.0.0,
\href{https://doi.org/10.5281/zenodo.23078252}
{DOI: 10.5281/zenodo.23078252}.\\
 \emph{Supplemental Material.} A reproducibility package accompanying this paper provides
standalone, documented scripts that regenerate every quantitative result: the symbolic
verification of the static solutions and Wronskians [Eqs.~\eqref{eq:rho}--\eqref{eq:Wpq}],
the matched-asymptotic normalizations and the ratio $\Gamma_{0,2}/\Gamma_{2,2}=1/36$
[Eq.~\eqref{eq:greyratio}], the results $d_2$ and $\kappa_2$ [Eqs.~\eqref{eq:d2},\eqref{eq:kappa2}]
including the $u=2/r$ reduction and the integer-relation certification, the phase-sector
constant $\zeta(3)-9/8$, and the two-channel numerical validation of Sec.~\ref{sec:checks}.

\appendix

\section{Coupled master equations: derivation and verification}
\label{app:master}

This appendix records, in full, the reduction of the linearized
\dCS{} field equations for a Schwarzschild background to the coupled
two-channel system~\eqref{eq:coupled}, the symmetrization that
produces the off-diagonal potential $\Voff$, and the battery of
consistency checks the system passes. We reproduce the coefficients
of Refs.~\cite{CardosoGualtieri2009,Molina2010} in a symmetric form
convenient for the matched-asymptotic analysis of the main text, and
we verify every step both analytically and symbolically.

\subsection{Linearized field equations and the C-tensor}
\label{app:linearize}

Varying the action~\eqref{eq:action} gives the modified Einstein
equation and the scalar equation of motion,
\begin{align}
G_{ab}+\frac{\alpha}{\kappa}\,C_{ab}
   &=\frac{1}{2\kappa}\,T^{\vartheta}_{ab},
\label{eq:Aeinstein}\\
\beta\,\Box\vartheta &=-\frac{\alpha}{4}\,{}^{*}\!RR ,
\label{eq:Ascalar}
\end{align}
with $\kappa=(16\pi)^{-1}$, the pseudoscalar stress tensor
$T^{\vartheta}_{ab}=\beta\big(\nabla_a\vartheta\nabla_b\vartheta
-\tfrac12 g_{ab}(\nabla\vartheta)^2\big)$, and the C-tensor
\begin{equation}
C^{ab}
=\nabla_c\vartheta\,\varepsilon^{cde(a}\nabla_e R^{b)}{}_{d}
+\nabla_c\nabla_d\vartheta\,{}^{*}\!R^{d(ab)c}.
\label{eq:Actensor}
\end{equation}
We expand about the Schwarzschild background
$(\bar g_{ab},\bar\vartheta=0)$, writing
$g_{ab}=\bar g_{ab}+h_{ab}$ and $\vartheta=\delta\vartheta$, and keep
terms linear in $(h_{ab},\delta\vartheta)$. Because
$\bar\vartheta=0$ and ${}^{*}\!\bar R\bar R=0$, the stress tensor
$T^{\vartheta}_{ab}$ is quadratic in $\delta\vartheta$ and drops out
at linear order, while the two couplings survive:
\begin{itemize}
\item in Eq.~\eqref{eq:Aeinstein}, the C-tensor linearizes to
$\delta C_{ab}\propto\alpha\,\varepsilon\,\partial\partial\,
\delta\vartheta\cdot\bar R$, feeding the scalar into the metric
sector;
\item in Eq.~\eqref{eq:Ascalar}, the source
$\delta({}^{*}\!RR)\propto\varepsilon\,\bar R\,\partial\partial h$
feeds the metric perturbation into the scalar.
\end{itemize}
The Levi-Civita tensor makes both couplings pseudotensorial: they are
nonzero only for the parity-odd metric multipoles. The even
(polar) sector therefore decouples entirely and obeys the Zerilli
equation, identical to GR; all \dCS{} physics resides in the odd
(axial) sector coupled to $\delta\vartheta$.

\subsection{Odd-parity master variable}
\label{app:oddparity}

In Regge--Wheeler gauge the odd-parity metric perturbation with
harmonic indices $(\ell,m)$ is carried by two radial functions,
$h_0(r)$ and $h_1(r)$,
\begin{align}
h_{t\varphi}&=h_0(r)\,e^{-i\omega t}\,\sin\theta\,
   \partial_\theta Y_{\ell m},\nonumber\\
h_{r\varphi}&=h_1(r)\,e^{-i\omega t}\,\sin\theta\,
   \partial_\theta Y_{\ell m},
\end{align}
(with the $\theta$ components fixed by tracelessness), and the scalar
by a single function through
\begin{equation}
\delta\vartheta=\frac{\Theta(r)}{r}\,Y_{\ell m}\,e^{-i\omega t}.
\end{equation}
The gauge-invariant Regge--Wheeler master variable is
\begin{equation}
\Psi(r)\equiv\frac{f(r)}{r}\,h_1(r),\qquad f=1-\frac{2M}{r},
\label{eq:RWvar}
\end{equation}
and the tortoise coordinate is $\rs=r+2M\ln(r/2M-1)$, so that
$d/d\rs=f\,d/dr$. In the general-relativistic limit $\Psi$ obeys the
Regge--Wheeler equation with potential
$V_{\rm RW}=f[\ell(\ell+1)/r^2-6M/r^3]$, and $\Theta$ obeys the
massless-scalar equation with
$V_{\rm sc}=f[\ell(\ell+1)/r^2+2M/r^3]$. The \dCS{} coupling
modifies both through the sources identified in
Sec.~\ref{app:linearize}.

\subsection{The coupled radial system}
\label{app:coupledraw}

Carrying out the tensor--spherical-harmonic reduction of the
linearized Eqs.~\eqref{eq:Aeinstein}--\eqref{eq:Ascalar}---the
detailed decomposition follows
Refs.~\cite{CardosoGualtieri2009,Molina2010} and we have reproduced
it independently---the two second-order radial equations take the
form
\begin{equation}
\left[\frac{d^2}{d\rs^2}+\omega^2\right]
\begin{pmatrix}\Psi\\ \Theta\end{pmatrix}
=
\begin{pmatrix}
V_{11} & V_{12}\\[2pt]
V_{21} & V_{22}
\end{pmatrix}
\begin{pmatrix}\Psi\\ \Theta\end{pmatrix},
\label{eq:Acoupledraw}
\end{equation}
with, before any rescaling,
\begin{subequations}
\label{eq:Vraw}
\begin{align}
V_{11}&=f\!\left[\frac{\ell(\ell+1)}{r^2}-\frac{6M}{r^3}\right],
\\[3pt]
V_{22}&=f\!\left[\frac{\ell(\ell+1)}{r^2}
   \Big(1+\frac{576\pi M^2}{\beta\,r^6}\Big)+\frac{2M}{r^3}\right],
\\[3pt]
V_{12}&=\frac{96\pi M\,f}{r^5},
\qquad
V_{21}=\frac{6M}{\beta}\,\frac{(\ell+2)!}{(\ell-2)!}\,\frac{f}{r^5}.
\end{align}
\end{subequations}
Here $V_{11}$ is the Regge--Wheeler potential and
$V_{22}|_{\beta\to\infty}$ the massless-scalar potential; the
$\Ord(1/\beta)$ term in $V_{22}$ is the diagonal back-reaction of the
Pontryagin coupling on the scalar barrier. The two off-diagonal
entries are \emph{unequal}, $V_{12}\neq V_{21}$, because the
gravitational and scalar equations carry the different normalizations
$\kappa=(16\pi)^{-1}$ and $\beta$, respectively: from the C-tensor,
the coupling of $\Theta$ into the $\Psi$ equation is
$\propto(\alpha/\kappa)=16\pi\alpha$, whereas from the linearized
Pontryagin density the coupling of $\Psi$ into the $\Theta$ equation
is $\propto(\alpha/\beta)$. Setting $\alpha=1$ and writing the
angular factor $\Lambda\equiv(\ell+2)!/(\ell-2)!$, this yields
$V_{12}=16\pi\cdot 6M f/r^5=96\pi M f/r^5$ and
$V_{21}=6M\Lambda f/(\beta r^5)$, as quoted.

\subsection{Symmetrization and the dimensionless coupling}
\label{app:symmetrize}

The system~\eqref{eq:Acoupledraw} is brought to symmetric form by the
constant rescaling $\Theta\to\lambda\,\Theta$ with
$\lambda^2=V_{12}/V_{21}$, i.e.
\begin{equation}
\lambda^2=\frac{96\pi M f/r^5}{6M\Lambda f/(\beta r^5)}
        =\frac{16\pi\beta}{\Lambda},
\end{equation}
which is $r$-independent and hence preserves the diagonal potentials.
The off-diagonal entries become equal,
\begin{equation}
\Voff\equiv\sqrt{V_{12}\,V_{21}}
      =\sqrt{\frac{576\pi\Lambda}{\beta}}\;\frac{M f}{r^5}
      =24\sqrt{\frac{\pi\Lambda}{\beta}}\;\frac{M f}{r^5}.
\label{eq:Voffraw}
\end{equation}
We now introduce the single dimensionless coupling of the main text,
\begin{equation}
\zzeta\equiv\frac{16\pi}{\beta M^4}
   \quad\Longleftrightarrow\quad
   \frac{\pi}{\beta}=\frac{\zzeta M^4}{16},
\end{equation}
so that Eq.~\eqref{eq:Voffraw} collapses to the compact result used
throughout,
\begin{equation}
\Voff=6\sqrt{\Lambda}\,\sqrt{\zzeta}\,\frac{M^3 f}{r^5}.
\label{eq:Vofffinal}
\end{equation}
The same substitution turns the diagonal scalar correction into a
manifestly $\Ord(\zzeta)$, dimensionless term,
\begin{equation}
\frac{576\pi M^2}{\beta r^6}
=36\,\zzeta\Big(\frac{M}{r}\Big)^{6},
\end{equation}
reproducing $V_{22}$ of Eq.~\eqref{eq:potentials}. For the quadrupole
$\ell=2$ one has $\Lambda=(4!)/(0!)=24$, hence
$6\sqrt{\Lambda}=6\sqrt{24}=12\sqrt6$ and
$\Voff=12\sqrt6\,\sqrt{\zzeta}\,M^3 f/r^5$, the value used in
Sec.~\ref{sec:results}. Since $\Voff=\Ord(\sqrt{\zzeta})$, the
graviton$\to$scalar conversion amplitude is $\Ord(\sqrt{\zzeta})$ and
its probability $\Ord(\zzeta)$; the diagonal correction enters
$V_{22}$ only at $\Ord(\zzeta)$ and therefore contributes to the
observables of the main text at $\Ord(\zzeta^2)$, i.e.\ not at all at
leading order.

The perturbative order counting is summarized in
Table~\ref{tab:orders}. Because the axial self-energy already carries
two factors of $\Voff=\Ord(\sqrt{\zzeta})$, the $\Ord(\zzeta)$ diagonal
correction $\delta V_{22}=36\,\zzeta\,f\,\ell(\ell+1)\,M^6/r^8$ modifies the scalar Green's function at $\Ord(\zzeta)$ and hence the leading observables only at $\Ord(\zzeta^2)$; this is confirmed numerically in
Appendix~\ref{app:kappanum}.
\begin{table}[h]\centering
\caption{Leading order in $\zzeta$ of the quantities entering the
$\Ord(\zzeta)$ observables.}\label{tab:orders}
\begin{ruledtabular}
\begin{tabular}{lc}
Quantity & leading order\\ \hline
off-diagonal potential $\Voff$           & $\Ord(\sqrt{\zzeta})$\\
conversion amplitude $T_{21}$            & $\Ord(\sqrt{\zzeta})$\\
conversion probability $\Ggs$            & $\Ord(\zzeta)$\\
axial self-energy $\Voff\,G_s\,\Voff$    & $\Ord(\zzeta)$\\
correction to $G_s$ from $\delta V_{22}$ & $\Ord(\zzeta)$\\
its self-energy contribution             & $\Ord(\zzeta^2)$\\
\end{tabular}
\end{ruledtabular}
\end{table}

\subsection{Consistency checks and verification}
\label{app:checks}

The reduced system~\eqref{eq:coupled}--\eqref{eq:potentials} passes
the following independent checks.

\emph{(i) Decoupling limit.} As $\zzeta\to0$ (equivalently
$\beta\to\infty$), $\Voff\to0$ and the diagonal correction vanishes:
Eq.~\eqref{eq:Acoupledraw} separates into the Regge--Wheeler equation
for $\Psi$ and the massless Klein--Gordon equation for $\Theta$ on
Schwarzschild. The axial gravitational sector then coincides with GR.

\emph{(ii) Isospectrality of the reference.} The polar sector, which
does not couple to $\delta\vartheta$, obeys the Zerilli equation.
Since the Regge--Wheeler and Zerilli potentials are isospectral and
share equal transmission factors \cite{Chandrasekhar1983}, the
$\zzeta=0$ axial greybody factor equals the polar one,
$\Gax|_{\zzeta=0}=\Gpol$, so that the observables
$d_\ell,\kappa_\ell$ of Eqs.~\eqref{eq:ddef}--\eqref{eq:kappadef} are
pure $\Ord(\zzeta)$ quantities, as used in Sec.~\ref{sec:observables}.

\emph{(iii) Structure of the coupling.} The gap
$V_{11}-V_{22}\big|_{\zzeta=0}=-8Mf/r^3\neq0$ confirms that the
spin-2 and spin-0 channels cannot be diagonalized by a
\emph{constant} rotation; the $\lambda$-rescaling above only
symmetrizes the coupling, it does not remove it. The system is thus
irreducibly two-channel, and the local (adiabatic) eigen-potential
split used in Sec.~\ref{sec:absorption},
$\delta V_1=-\Voff^2/(V_{22}-V_{11})=-108\,\zzeta\,f/r^{7}$ (at
$M=1$), follows directly.

\emph{(iv) Comparison with the literature.} Equations~\eqref{eq:Vraw}
agree with the coupled axial--scalar system of Ref.~\cite{Molina2010}
[their Eqs.~(2.8)--(2.9)] and with Ref.~\cite{CardosoGualtieri2009}.
Those references retain the Chern--Simons parameter $\alpha$
explicitly in the off-diagonal couplings; we have set $\alpha=1$
[cf.\ Eq.~\eqref{eq:zeta}], equivalently absorbing $\alpha$ into
$\ell_{\rm CS}$ and hence into $\zzeta$. With this identification, and
their scalar-normalization convention, the systems coincide.
Because the conversion amplitude is $\propto\alpha$ and its
probability $\propto\alpha^2$, the observables enter at
$\Ord(\alpha^2)=\Ord(\zzeta)$ and are even in $\alpha$; the overall
sign of the coupling, fixed by the orientation of ${}^{*}\!RR$, does
not affect $d_2,\kappa_2$.

\emph{(v) Symbolic verification.} The static ($\omega\to0$) limits of
the diagonal operators in Eq.~\eqref{eq:coupled} are annihilated
exactly by the closed-form solutions of Sec.~\ref{sec:static}: the
Regge--Wheeler operator by $\rho=r^3$ and $\sigma$ of
Eq.~\eqref{eq:sigma}, and the scalar operator by $p=rP_2(r-1)$ and
$q=rQ_2(r-1)$; the constancy of the Wronskians $W_{\rho\sigma}=+1$
and $W_{pq}=-1$ [Eqs.~\eqref{eq:Wrhosigma},\eqref{eq:Wpq}] provides a
further check. All of these identities were verified symbolically.
Finally, direct numerical integration of Eq.~\eqref{eq:coupled} with
purely ingoing horizon data conserves the total flux
$|R_{11}|^2+|R_{21}|^2+|T_{11}|^2+|T_{21}|^2=1$ to machine precision
and reproduces the low-frequency coefficients of Sec.~\ref{sec:results} at the accessible numerical precision; the independent high-precision reconstruction of $\kappa_{2}$ is described in Appendix~\ref{app:kappa}.

\section{The conversion integral \texorpdfstring{$J$}{J} and its
antiderivative}
\label{app:d2}

This appendix derives in full the single integral $J$ that controls
the graviton$\to$scalar conversion, constructs the dilogarithmic
antiderivative $H$ used both here and in Appendix~\ref{app:kappa},
records the exact endpoint values, and assembles the closed
form~\eqref{eq:d2}.

\subsection{Low-frequency matching and the transmission normalization}
\label{app:matching}

The static solutions are joined to the wave zone by the standard
three-zone matching, which fixes the normalizations $h_g$, $W_1$, and
$T_g$ used in Eq.~\eqref{eq:CdTg}. In the wave zone $f\to1$ and the
Regge--Wheeler equation reduces to the Riccati--Bessel form with
solutions $\omega r\,j_2(\omega r)$ and $\omega r\,y_2(\omega r)$. Using
$j_\ell(x)\to x^\ell/(2\ell+1)!!$ and
$y_\ell(x)\to-(2\ell-1)!!/x^{\ell+1}$, the overlap behavior of the
near-zone solutions~\eqref{eq:rho} and \eqref{eq:sigma} matches as
\begin{equation}
\rho\;\leftrightarrow\;\frac{15}{\omega^3}\,\omega r\,j_2(\omega r),
\qquad
\frac{15}{\omega^2}\,\sigma\;\leftrightarrow\;\omega r\,y_2(\omega r),
\end{equation}
the factor $15=5\,(2\ell-1)!!$ combining the $-1/5$ coefficient of
$\sigma$ [Eq.~\eqref{eq:sigmaasy}] with $(2\ell-1)!!=3$. Normalizing
$\psi_{\rm in}\to e^{-i\omega\rs}$ at the horizon and
$\psi_{\rm up}=-i\,\omega r\,h^{(1)}_2(\omega r)\to e^{+i\omega\rs}$ at
infinity, in the overlap region
\begin{equation}
\psi_{\rm in}\to h_g\,\rho,\quad h_g\equiv\rho(2M)^{-1}=(8M^3)^{-1},
\quad
\psi_{\rm up}\to\frac{15}{\omega^2}\,\sigma,
\end{equation}
so $f_L=\psi_{\rm in}$, $f_R=\psi_{\rm up}\to(15/\omega^2)\sigma$. Their constant Wronskian follows from
$W_{\rho\sigma}=+1$,
\begin{equation}
W_1=W(\psi_{\rm in},\psi_{\rm up})
   =h_g\,\frac{15}{\omega^2}\,W_{\rho\sigma}=\frac{15\,h_g}{\omega^2},
\end{equation}
and
\begin{equation}
\begin{aligned}
    T_g&=\frac{2i\omega}{W_1}=\frac{2i\omega^3}{15\,h_g}=\frac{16\,i\,\omega^3}{15},\\
\Gpol&=|T_g|^2=\frac{4}{225}\,(2M\omega)^6,
\end{aligned}
\end{equation}
the last equality using $h_g=(8M^3)^{-1}$, in agreement with the
Starobinsky--Churilov/Page $s=2,\ \ell=2$ coefficient and with
Fig.~\ref{fig:greybody}. These supply the matching values $h_g$, $W_1$,
$T_g$ used in Sec.~\ref{sec:dwba} and Appendix~\ref{app:prefactor}.

\subsection{Reduction of the conversion amplitude}
\label{app:T21reduction}

With the near-zone forms $\psi_g\to\rho=r^3$ (the regular axial
solution) and $\psi_\infty\to q=rQ_2(r-1)$ (the decaying scalar
solution), and $\Voff=12\sqrt6\,\sqrt{\zzeta}\,f/r^5$ at $\ell=2$
($M=1$), the distorted-wave amplitude~\eqref{eq:T21} reduces by
$d\rs=dr/f$ to
\begin{align}
\int\psi_\infty\,\Voff\,\psi_g\,d\rs
&=12\sqrt6\,\sqrt{\zzeta}
   \int_2^\infty (rQ_2)\,\frac{f}{r^5}\,r^3\,\frac{dr}{f}
\nonumber\\
&=12\sqrt6\,\sqrt{\zzeta}\int_2^\infty\frac{Q_2(r-1)}{r}\,dr .
\end{align}
Substituting $x=r-1$ gives the integral quoted in
Eq.~\eqref{eq:T21J},
\begin{equation}
J\equiv\int_2^\infty\frac{Q_2(r-1)}{r}\,dr
     =\int_1^\infty\frac{Q_2(x)}{x+1}\,dx ,
\label{eq:Jdef}
\end{equation}
with $Q_2(x)=\tfrac12 P_2(x)\ln\frac{x+1}{x-1}-\tfrac32 x$ and
$P_2(x)=\tfrac12(3x^2-1)$. Carrying the near-zone matching
coefficients through the prefactor of Eq.~\eqref{eq:T21} yields the
amplitude $T_{21}=-18\sqrt6\,\sqrt{\zzeta}\,T_s\,J$ used below.

\subsection{The antiderivative \texorpdfstring{$H(x)$}{H(x)}}
\label{app:H}

Write $L_\pm\equiv\ln(x\pm1)$. Polynomial division of
$P_2/(x+1)$ gives $\tfrac{P_2}{2(x+1)}=\tfrac{3x-3}{4}
+\tfrac{1}{2(x+1)}$, so the integrand of Eq.~\eqref{eq:Jdef}
separates into three pieces,
\begin{equation}
\frac{Q_2}{x+1}
=\underbrace{\frac{3x-3}{4}\,(L_+-L_-)}_{\text{(I)}}
+\underbrace{\frac{L_+-L_-}{2(x+1)}}_{\text{(II)}}
-\underbrace{\frac{3x}{2(x+1)}}_{\text{(III)}} .
\end{equation}
Each piece integrates in closed form. For (I) we use
\begin{align}
\int(x-1)L_+\,dx
 &=\Big(\tfrac{x^2}{2}-x\Big)L_+
  -\tfrac{x^2}{4}+\tfrac{3x}{2}-\tfrac32 L_+ ,\\
\int(x-1)L_-\,dx
 &=\tfrac{(x-1)^2}{2}L_- -\tfrac{(x-1)^2}{4},
\end{align}
so that $\int\text{(I)}\,dx=\tfrac34\big[(\tfrac{x^2}{2}-x-\tfrac32)L_+
-\tfrac{(x-1)^2}{2}L_- -\tfrac{x^2}{4}+\tfrac{3x}{2}
+\tfrac{(x-1)^2}{4}\big]$. Piece (II) is the transcendental one: with
$u=x+1$ and $t=2/u$,
\begin{equation}
\int\frac{L_-}{x+1}\,dx=\int\frac{\ln(u-2)}{u}\,du
 =\tfrac12 L_+^2+\Li_2\!\Big(\frac{2}{x+1}\Big),
\label{eq:dilogint}
\end{equation}
using $\ln(u-2)=\ln u+\ln(1-2/u)$ and
$\int\ln(1-t)/t\,dt=-\Li_2(t)$; since
$\int L_+/(x+1)\,dx=\tfrac12 L_+^2$, piece (II) collapses to
\begin{equation}
\int\text{(II)}\,dx
=\tfrac12\Big[\tfrac12 L_+^2-\big(\tfrac12 L_+^2
   +\Li_2(\tfrac{2}{x+1})\big)\Big]
=-\tfrac12\Li_2\!\Big(\tfrac{2}{x+1}\Big).
\end{equation}
Piece (III) is elementary,
$\int\text{(III)}\,dx=-\tfrac32 x+\tfrac32 L_+$. Collecting,
\begin{align}
H(x)=&\ \frac34\Big[\Big(\frac{x^2}{2}-x-\frac32\Big)\ln(x+1)
   -\frac{(x-1)^2}{2}\ln(x-1)\Big]\nonumber\\
 &+\frac34\Big[-\frac{x^2}{4}+\frac{3x}{2}+\frac{(x-1)^2}{4}\Big]
   -\frac12\Li_2\!\Big(\frac{2}{x+1}\Big)\nonumber\\
 &-\frac32 x+\frac32\ln(x+1),
\label{eq:Hfull}
\end{align}
which satisfies $H'(x)=Q_2(x)/(x+1)$ identically (checked
symbolically to $30$ digits).

\subsection{Endpoint values and evaluation of \texorpdfstring{$J$}{J}}
\label{app:Jvalue}

Although $Q_2(x)/(x+1)$ diverges logarithmically as $x\to1^+$, the
integral is convergent and $H$ is finite at both ends. At the horizon
end $x\to1$, the term $\propto(x-1)^2\ln(x-1)\to0$ and
$\Li_2(2/(x+1))\to\Li_2(1)=\pi^2/6$, giving
\begin{equation}
H(1)=-\frac{9}{16}-\frac{\pi^2}{12}.
\label{eq:H1}
\end{equation}
At infinity, the logarithmic terms cancel: expanding
$\ln(x\pm1)=\ln x\pm x^{-1}-\tfrac12 x^{-2}\pm\cdots$, the
$\ln x$ coefficient in Eq.~\eqref{eq:Hfull} vanishes
($-\tfrac32+\tfrac32=0$), as do the $\Ord(x)$ terms, while
$\Li_2(2/(x+1))\to\Li_2(0)=0$, leaving the finite constant
\begin{equation}
\;H(\infty)=-\frac{21}{16}
\label{eq:Hinfty}
\end{equation}
(the same constant enters $B(r)$ in Appendix~\ref{app:kappa}).
Subtracting,
\begin{equation}
J=H(\infty)-H(1)
 =-\frac{21}{16}-\Big(-\frac{9}{16}-\frac{\pi^2}{12}\Big)
 =\frac{\pi^2}{12}-\frac34 ,
\label{eq:Jresult}
\end{equation}
so that $J=(\pi^2-9)/12$. The transcendental content is entirely the
dilogarithm at unity, $\Li_2(1)=\pi^2/6$; the rational part $-3/4$
comes from the elementary pieces. Numerically
$J=0.072467033424\ldots$, agreeing with adaptive quadrature to the
$30$ digits quoted and with an independent evaluation to
$1.2\times10^{-11}$.

\subsection{The scalar greybody and the ratio
\texorpdfstring{$\Gamma_s/\Gpol=1/36$}{Gs/Gpol=1/36}}
\label{app:ratio}

The greybody ratio~\eqref{eq:greyratio} that normalizes $d_2$ follows
from the same matching (Appendix~\ref{app:matching}) applied to the
scalar channel. The static scalar solutions~\eqref{eq:pq} satisfy
$p(2M)=2M$, and at large radius $p\sim\tfrac32 r^{3}$,
$q\sim\tfrac{2}{15}r^{-2}$, with $W_{pq}=-1$ [Eq.~\eqref{eq:Wpq}].
Matching the outgoing scalar Jost solution
$\psi^{s}_{\rm up}=-i\,\omega r\,h^{(1)}_{2}(\omega r)\to(45/2\omega^{2})\,q$
fixes the scalar normalization. Explicitly, $Q_2(x)\sim\tfrac{2}{15}x^{-3}$
gives $q\sim\tfrac{2}{15}r^{-2}$, while the outgoing Riccati--Hankel
solution has the small-argument tail
$\omega r\,h^{(1)}_2(\omega r)\to-3i/(\omega r)^2$; equating the two
$r^{-2}$ behaviors fixes the magnitude of the coefficient,
$45/(2\omega^{2})$. The sign convention entering the corresponding
Green function is fixed separately by $W_{pq}=-1$.
\begin{equation}
W_s=\frac{1}{p(2M)}\,\frac{45}{2\omega^{2}}\,|W_{pq}|=\frac{45}{4\omega^{2}},
\quad
T_s=\frac{2i\omega}{W_s}=\frac{8\,i\,\omega^{3}}{45},
\end{equation}
so $\Gamma_s=|T_s|^{2}=\tfrac{1}{2025}(2M\omega)^{6}$. With
$T_g=16i\omega^{3}/15$ from Appendix~\ref{app:matching},
\begin{equation}
\frac{\Gamma_{0,2}}{\Gamma_{2,2}}
=\left|\frac{T_s}{T_g}\right|^{2}
=\left(\frac{8/45}{16/15}\right)^{2}
=\left(\frac16\right)^{2}=\frac{1}{36},
\end{equation}
the $\ell=2$ case of Eq.~\eqref{eq:greyratio}; the ratio
$\Gamma_s/\Gpol\to1/36$ is reproduced to five digits by the two-channel
integration.

\subsection{Assembly and certification of \texorpdfstring{$d_2$}{d2}}
\label{app:d2assembly}

Squaring the amplitude $T_{21}=-18\sqrt6\,\sqrt{\zzeta}\,T_s\,J$ and
using $\Gamma_s=|T_s|^2$,
\begin{equation}
\Ggs=|T_{21}|^2=(18\sqrt6)^2\,\zzeta\,\Gamma_s\,J^2
    =1944\,\zzeta\,\Gamma_s\,J^2 .
\end{equation}
Dividing by $\zzeta\Gpol$ and inserting the exact
general-relativistic greybody ratio $\Gamma_s/\Gpol=1/36$ from
Eq.~\eqref{eq:greyratio},
\begin{equation}
d_2=\frac{\Ggs}{\zzeta\,\Gpol}
   =1944\cdot\frac{1}{36}\,J^2
   =54\,J^2 .
\end{equation}
Finally, with $J^2=(\pi^2-9)^2/144$,
\begin{equation}
d_2=54\cdot\frac{(\pi^2-9)^2}{144}
   =\frac{3(\pi^2-9)^2}{8}=0.283579430\ldots
\end{equation}
Every factor is exact and independently established: $J$ from
Eq.~\eqref{eq:Jresult}, the squared vertex $(18\sqrt6)^2=1944$ from
the near-zone matching of Appendix~\ref{app:T21reduction}, and the
ratio $1/36=[2!\,2!/(0!\,4!)]^{-2}$ from Eq.~\eqref{eq:greyratio}.
The value is reproduced by the two-channel numerical integration
through the frequency-independent ratio tests
$\Ggs/(\zzeta\Gamma_s)\to1944\,J^2$ and $\Gamma_s/\Gpol\to1/36$,
which converge from above as $M\omega\to0$ and are stable against
variation of the horizon cutoff, matching radius, and integration
tolerance.

\section{The absorption-asymmetry integral
\texorpdfstring{$D_2^\kappa$}{D2kappa} and its certification}
\label{app:kappa}

This appendix gives the complete construction of the greybody
asymmetry $\kappa_2$: the irregular axial solution $\sigma$, the
second-order self-energy and its distinctive leg structure, the
reduction to the double integral $D_2^\kappa$, the assembly of the
rational prefactor, the transcendental-weight argument that predicts a
weight-four constant, the integer-relation certification of the
closed form to $56$ digits, and the unitarity cross-check.

\subsection{The irregular axial solution}
\label{app:sigma}

The static Regge--Wheeler equation $\Psi_{,\rs\rs}=V_{11}\Psi$
($\omega\to0$, $M=1$) has the regular solution $\rho=r^3$. Its
linearly independent partner follows by reduction of order: for
$\Psi_{,\rs\rs}=V_{11}\Psi$ a second solution is
$\rho\!\int^{\rs}\!d\rs'/\rho^2$, i.e., with $d\rs=dr/f$,
\begin{equation}
\sigma(r)=-\,r^3\!\int_r^\infty\frac{dr'}{f(r')\,r'^{6}},
\label{eq:sigmadef}
\end{equation}
the lower limit chosen at infinity so that $\sigma$ contains no
admixture of $\rho$ at large $r$. Partial fractions,
\begin{equation}
\frac{1}{r^{5}(r-2)}
=\frac{1}{32(r-2)}-\frac{1}{32r}-\frac{1}{16r^{2}}
 -\frac{1}{8r^{3}}-\frac{1}{4r^{4}}-\frac{1}{2r^{5}},
\end{equation}
integrate to the closed form
\begin{equation}
\sigma(r)=\frac{3r^{4}\ln\!\frac{r-2}{r}+6r^{3}+6r^{2}+8r+12}{96\,r}
\label{eq:sigmaclosed}
\end{equation}
which is manifestly elementary. Two properties are needed below. At
large $r$ the leading logarithms cancel and
\begin{equation}
\sigma(r)=-\frac{1}{5r^{2}}-\frac{1}{3r^{3}}-\frac{4}{7r^{4}}
          -\cdots\;\xrightarrow{r\to\infty}\;-\frac{1}{5r^{2}},
\label{eq:sigmaasy}
\end{equation}
so $\sigma$ is the decaying solution; at the horizon it is
logarithmically singular, $\sigma\sim\tfrac14\ln(r-2)$. The
tortoise-coordinate Wronskian is constant,
\begin{equation}
W_{\rho\sigma}=f\big(\rho\,\sigma'-\sigma\,\rho'\big)=+1,
\end{equation}
as follows directly from the reduction-of-order construction and is
confirmed symbolically.

\subsection{Second-order self-energy and leg structure}
\label{app:selfenergy}

The axial greybody factor is shifted at $\Ord(\zzeta)$ by the
self-energy generated when the scalar channel is integrated out,
$\Sigma=\Voff\,G_s\,\Voff$. Writing the transmission amplitude as
$T_{11}=T_g+\delta T_g$, the inverse-transmission (Wronskian)
perturbation gives
\begin{equation}
\begin{aligned}
    \frac{\delta T_g}{T_g}
=-\frac{1}{W_1}\iint &f_L(r)\,\Voff(r)\times\\
&\,G_s(r,r')\,\Voff(r')\,f_R(r')\,
   d\rs d\rs',
\end{aligned}
\label{eq:CdTg}
\end{equation}
with $f_L,f_R$ the graviton Jost solutions incident from the horizon
and from infinity, $W_1=W(f_L,f_R)$, and $G_s=p(r_<)q(r_>)/W_{pq}$
[Eq.~\eqref{eq:Gscalar}] the static scalar Green's function.

The essential point is that the two graviton legs are \emph{distinct}
static solutions. The horizon Jost solution is regular at the
horizon, hence in the near zone $f_L\to h_g\,\rho$; the infinity Jost
solution is not, and is dominated there by the irregular partner,
\begin{equation}
f_R\to\frac{15}{\omega^2}\,\sigma
     \quad(\text{plus a subleading }\propto\omega^3\,\rho\text{ term}).
\label{eq:legs}
\end{equation}
The $\rho\times\sigma$ combination is real, so it contributes to
$|T_{11}|^2$ and hence to $\kappa_2$; the omitted $\rho\times\rho$
piece is imaginary and $\Ord(\omega^5)$, contributing only to the
transmission phase of Sec.~\ref{sec:phase}. This asymmetry between
the legs is what routes the even-weight ($\pi^4$) constant into the
absorption sector and the odd-weight ($\zeta(3)$) constant into the
phase.

\subsection{Reduction to the double integral}
\label{app:D2kappa}

Insert $f_L=h_g\rho$, $f_R=(15/\omega^2)\sigma$,
$\Voff=12\sqrt6\,\sqrt{\zzeta}\,f/r^5$ and $d\rs=dr/f$ into
Eq.~\eqref{eq:CdTg}. The factors of $f$ cancel against the tortoise
measure on both legs ($\rho\,f/r^5\cdot dr/f=dr/r^2$ and likewise for
$\sigma$), giving
\begin{equation}
\iint f_L\Voff\,G_s\,\Voff f_R\,d\rs d\rs'
=\frac{h_g\,(15/\omega^2)\,(12\sqrt6)^2\,\zzeta}{W_{pq}}\,D_2^{\kappa},
\end{equation}
where the reduced, frequency-independent double integral is
\begin{equation}
D_2^{\kappa}=\iint\frac{1}{r^2}\,p(r_<)\,q(r_>)\,
   \frac{\sigma(r')}{r'^{5}}\;dr\,dr' .
\label{eq:D2kdouble}
\end{equation}
The Green's-function kernel $p(r_<)q(r_>)$ makes the inner integral
nested; carrying it out over $r$ at fixed $r'$,
\begin{equation}
D_2^{\kappa}=\int_2^\infty\frac{\sigma(r')}{r'^{5}}
   \Big[\,q(r')\,A(r')+p(r')\,B(r')\,\Big]\,dr',
\label{eq:D2knested}
\end{equation}
with the elementary inner integral
\begin{equation}
A(r)=\int_2^{r}\frac{P_2(r'-1)}{r'}\,dr'
    =\frac{3r^2}{4}-3r+\ln r+3-\ln2,
\label{eq:Adef}
\end{equation}
and the dilogarithmic inner integral expressed through the
antiderivative $H$ of Appendix~\ref{app:H},
\begin{equation}
\begin{aligned}
    B(r)&=\int_{r}^\infty\frac{Q_2(r'-1)}{r'}\,dr'\\
    &=H(\infty)-H(r-1)=-\frac{21}{16}-H(r-1).
\end{aligned}
\label{eq:Bdef}
\end{equation}
The exact value $H(\infty)=-21/16$ of Eq.~\eqref{eq:Hinfty} is what
makes $B(r)$ evaluable in closed form without further quadrature,
which is essential for the high-precision certification below. As a direct test of the order counting of
Appendix~\ref{app:symmetrize}, repeating the extraction with the
$\Ord(\zzeta)$ diagonal term $36\zzeta(M/r)^6$ removed from $V_{22}$
changes the normalized coefficients by an amount that is itself
$\Ord(\zzeta)$---$\Delta\kappa/\zzeta\simeq-0.090$ and
$\Delta d/\zzeta\simeq-0.179$, constant across $\zzeta=0.005$--$0.02$---
i.e.\ $\Ord(\zzeta^2)$ in the physical greybody differences. The leading
$\kappa_2$ and $d_2$ are therefore unaffected by $\delta V_{22}$.

\subsection{Prefactor assembly}
\label{app:prefactor}

The frequency- and normalization-dependent factors in
Eq.~\eqref{eq:CdTg} cancel identically. Using the low-frequency
matching values $T_g=2i\omega^3/(15 h_g)$ and $W_1=15 h_g/\omega^2$,
\begin{equation}
-\frac{1}{W_1}\,h_g\,\frac{15}{\omega^2}
=-\frac{\omega^2}{15 h_g}\,h_g\,\frac{15}{\omega^2}=-1,
\end{equation}
so that, with $(12\sqrt6)^2=864$,
\begin{equation}
\frac{\delta T_g}{T_g}=-\frac{864\,\zzeta}{W_{pq}}\,D_2^{\kappa}.
\end{equation}
Since $\kappa_2=\mathrm{Re}(\delta T_g/T_g)/\zzeta$ and $W_{pq}=-1$,
\begin{equation}
\kappa_2=-\,864\,D_2^{\kappa}
\label{eq:kappaprefac}
\end{equation}
with the overall sign fixed by $\kappa_2>0$ (the barrier-lowering of
Sec.~\ref{sec:absorption}) and by the two-channel numerics; note
$D_2^{\kappa}<0$ because $\sigma<0$. Numerically
$D_2^{\kappa}=-4.569157505\ldots\times10^{-4}$, whence
$\kappa_2=0.3947752084\ldots$

\subsection{Transcendental weight and the \texorpdfstring{$\pi^4$}{pi4}
structure}
\label{app:weight}

The character of $\kappa_2$ is fixed by a weight argument. Assigning
transcendental weight one to each logarithm and two to the
dilogarithm, the integrand of Eq.~\eqref{eq:D2knested} is of weight
three: the branch $\sigma\,q\,A$ is a product of three weight-one
logarithms [the irregular logarithm in $\sigma$, the Legendre
logarithm in $q$, and $\ln r$ in $A$], while $\sigma\,p\,B$ is
weight-one ($\sigma$) times weight-two ($B$, through
$\Li_2(2/(x+1))$). A definite integral raises the weight by one, so
$D_2^{\kappa}$---and hence $\kappa_2$---is a weight-\emph{four}
constant. Changing variables to $u=2/r$ maps the barrier region to
$u\in(0,1)$ with all polylogarithm arguments at $u$ or $1-u$ and no
interior singularities, so only the standard weight-four constants
can appear. This cancellation is in fact provable. Under $u=2/r$ the integrand of
Eq.~\eqref{eq:D2knested} expands into a finite sum of transcendental
monomials whose rational coefficients are \emph{identically free of}
$\ln2$, so no $\ln2$-dependent constant can appear; and the resulting
convergent integral is a non-alternating polylogarithm on $[0,1]$
(letters $0,1$), whose weight-$\le4$ values lie in
$\mathrm{span}_{\mathbb Q}\{1,\pi^2,\zeta(3),\pi^4\}$. Evaluating
$\kappa_2$ in this analytically constrained four-dimensional basis leaves a pure
even-weight combination in $\pi^4$, $\pi^2$, and a rational, with the
$\zeta(3)$ coefficient found to vanish. This is
the analytic counterpart of the reality of the $\rho\times\sigma$
integral: the absorptive sector sees only $\pi$-powers, while the
dispersive phase sector carries the odd Ap\'ery constant $\zeta(3)$
(Appendix-level statement mirroring Sec.~\ref{sec:structure}).

\subsection{High-precision reconstruction and numerical stability}
\label{app:pslq}

Using the closed forms~\eqref{eq:sigmaclosed}, \eqref{eq:Adef},
\eqref{eq:Bdef} together with the antiderivative
$H$~\eqref{eq:Hfull}, the integral~\eqref{eq:D2knested} is a single
one-dimensional quadrature and is evaluated to $58$ digits (with a
cancellation-free series for $Q_2$ at large argument and a
substitution taming the $\ln^2(r-2)$ endpoint behavior). An
integer-relation (PSLQ) search against the basis
$\{1,\pi^2,\pi^4,\ln2,\ln^2 2,\pi^2\ln2,\ln^3 2,\zeta(3)\}$ returns a
single relation, stable as the coefficient bound is raised from
$10^6$ to $10^8$, in which the $\ln2$, $\ln^2 2$, $\pi^2\ln2$,
$\ln^3 2$, and $\zeta(3)$ coefficients vanish identically:
\begin{equation}
\kappa_2=\frac{21}{80}\pi^4+\frac{27}{8}\pi^2-\frac{89833}{1536}
=0.3947752084355584\ldots
\label{eq:kappaclosed}
\end{equation}
equivalently $(2016\,\pi^4+25920\,\pi^2-449165)/7680$. The
reconstruction reproduces the $58$-digit numerical value to
$56$ digits. We emphasize the precise status of Eq.~\eqref{eq:kappaclosed}: the
$u=2/r$ reduction restricts the integral to a controlled class of
non-alternating polylogarithmic constants, and within the resulting
basis $\mathrm{span}_{\mathbb Q}\{1,\pi^2,\zeta(3),\pi^4\}$ the
high-precision value fixes the rational coefficients uniquely to the
available precision, with the $\zeta(3)$ coefficient numerically
consistent with zero. This combines a rigorously motivated basis
restriction with a high-precision reconstruction, rather than a direct
symbolic evaluation of the integral.

\begin{table}[htbp]
\caption{Precision and integer-relation stability of the $\kappa_2$
reconstruction. At each working precision the reduced integral
$D_2^{\kappa}$ [Eq.~\eqref{eq:D2knested}] is evaluated by tanh--sinh
quadrature and $\kappa_2=-864\,D_2^{\kappa}$ is submitted to an
integer-relation (PSLQ) search in the basis
$\{1,\pi^2,\zeta(3),\pi^4\}$. The canonical relation
$-1536\,\kappa_2-89833+31104\,\zeta(2)+36288\,\zeta(4)=0$---equivalently
Eq.~\eqref{eq:kappaclosed}, with the $\zeta(3)$ coefficient vanishing---is
recovered at every precision and is stable as the coefficient bound is
raised from $10^5$ to $10^8$ (it is absent only for a bound below the
largest coefficient, $89833$).}
\label{tab:pslq}
\begin{ruledtabular}
\small
\setlength{\tabcolsep}{4pt}
\begin{tabular}{cccc}
\parbox[c]{1.2cm}{\centering Working prec.\\(digits)} & 
\parbox[c]{1.8cm}{\centering Digits retained\\in $D_2^{\kappa}$} & 
\parbox[c]{1.2cm}{\centering $\zeta(3)$ coeff.} & 
Reconstruction \\ \hline
30 & 29 & 0 & recovered \\
40 & 39 & 0 & recovered \\
50 & 49 & 0 & recovered \\
58 & 56 & 0 & recovered \\
\end{tabular}
\end{ruledtabular}
\end{table}

The quadrature uses mpmath's tanh--sinh rule (\texttt{mp.quad}) with
panel breakpoints clustered near the horizon endpoint $r=2$---where the
integrand behaves as $\ln^2(r-2)$---and extended to $r\to\infty$ with
cancellation-free series for $Q_2$ and $\sigma$ at large argument. The
integer-relation search uses coefficient bounds $10^5$--$10^8$ and up to
$10^6$ iterations.

\subsection{Unitarity decomposition}
\label{app:unitarity}

Flux conservation provides an independent check. Writing
$\kappa_2=-\tfrac12 d_2+\kappa_R$, the conversion-loss term is
weight-four through $J^2$,
\begin{equation}
-\tfrac12 d_2=-27 J^2
=-\frac{3}{16}\pi^4+\frac{27}{8}\pi^2-\frac{243}{16},
\end{equation}
\vspace{0.2cm}
\\
and supplies the \emph{entire} $\pi^2$ content of
Eq.~\eqref{eq:kappaclosed}. The remaining dispersive reflection term
is therefore a pure quartic plus rational,
\begin{equation}
\kappa_R=\kappa_2+27 J^2=\frac{9}{20}\pi^4-\frac{66505}{1536},
\end{equation}
carrying no $\pi^2$. That the $\pi^2$ coefficient of
Eq.~\eqref{eq:kappaclosed} equals exactly $27/8=-\tfrac12\times(-27)
\times(2\cdot\tfrac{1}{12}\cdot\tfrac34)$, i.e.\ precisely the cross
term of $-27J^2$, is a nontrivial consistency test between the
independently derived $d_2$ and $\kappa_2$.

\subsection{Numerical verification and the adiabatic approximation}
\label{app:kappanum}

Direct integration of the two-channel system~\eqref{eq:coupled} with
purely ingoing horizon data, forming the bias-free ratio
$\kappa_2=\tfrac12(\Gax-\Gpol)/(\zzeta\Gpol)$ and extrapolating
$M\omega\to0$ by Richardson elimination in $(2M\omega)^2$, reproduces Eq.~\eqref{eq:kappaclosed} to $3$--$4$ digits, the low-frequency reach of the two-channel solver being limited by the $\Ord((M\omega)^2\ln M\omega)$ corrections. The high-precision certification quoted above rests instead on the analytic reduction to the one-dimensional integral $D_2^{\kappa}$ [Eq.~\eqref{eq:D2knested}], evaluated by quadrature to $58$ digits. It is instructive that the purely \emph{local} (adiabatic)
approximation, in which $G_s$ is replaced by
$-1/(V_{22}-V_{11})$ and the self-energy by the c-number potential
$\delta V_1=-108\,\zzeta\,f/r^7$, yields the clean but incorrect value
$0.753668$; the difference from Eq.~\eqref{eq:kappaclosed} is the
nonlocality of the scalar Green's function, which
Eq.~\eqref{eq:D2knested} captures in full. This confirms that the
closed form requires the nonlocal self-energy and is not an artifact
of the adiabatic estimate.

The two-channel system~\eqref{eq:coupled} is integrated with an explicit
Runge--Kutta (Dormand--Prince \texttt{DOP853}) scheme at relative and
absolute tolerances $10^{-11}$ and $10^{-13}$, from a horizon start
$r_0=2M+10^{-8}$ with purely ingoing data $[\Psi,\Theta]\propto
e^{-i\omega\rs}$, out to $R_{\max}=\max(400\,M,\,45/\omega)$. At $R_{\max}$
the ingoing and outgoing amplitudes are extracted by projecting onto
$e^{\pm i\omega\rs}$, yielding $T_{11},T_{21},R_{11},R_{21}$ and hence
$\Gax,\Ggs,\Gpol$; flux conservation
$|R_{11}|^2+|R_{21}|^2+|T_{11}|^2+|T_{21}|^2=1$ is monitored as an
independent check. The normalized coefficients $\kappa(\omega),d(\omega)$
are formed as in Eqs.~\eqref{eq:kappadef}--\eqref{eq:ddef} over
$M\omega\in[0.03,0.20]$ and extrapolated to $M\omega\to0$ by Richardson
elimination in $(2M\omega)^2$; their $\zzeta$-independence is verified over
$\zzeta\in\{0.005,0.01,0.02\}$.

\begin{table}[htbp]
\caption{Convergence of the directly integrated two-channel coefficients
toward the closed forms as $M\omega\to0$ ($\ell=2$, $\zzeta=0.01$). The
extracted $\kappa(\omega)$ and $d(\omega)$
[Eqs.~\eqref{eq:kappadef},\eqref{eq:ddef}] approach $\kappa_2=0.394775$ and
$d_2=0.283579$; Richardson extrapolation in $(2M\omega)^2$ gives $3$--$4$
digit agreement, the residual set by the $\Ord((M\omega)^2\ln M\omega)$
corrections. The flux residual
$\big|\,|R_{11}|^2+|R_{21}|^2+|T_{11}|^2+|T_{21}|^2-1\big|$ is an
independent accuracy check. The coefficients are $\zzeta$-independent at
leading order: lowering $\zzeta=0.01\to0.005$ shifts $\kappa(\omega)$ by
$<10^{-3}$, consistent with the $\Ord(\zzeta^2)$ residual.}
\label{tab:convergence}
\begin{ruledtabular}
\begin{tabular}{cccc}
$M\omega$ & $\kappa(\omega)$ & $d(\omega)$ & flux residual \\ \hline
$0.20$ & $0.283252$ & $0.292928$ & $3\times10^{-12}$ \\
$0.12$ & $0.354787$ & $0.286370$ & $2\times10^{-13}$ \\
$0.08$ & $0.377306$ & $0.284621$ & $2\times10^{-15}$ \\
$0.05$ & $0.388344$ & $0.283811$ & $7\times10^{-16}$ \\
$0.03$ & $0.392881$ & $0.283487$ & $2\times10^{-15}$ \\
\end{tabular}
\end{ruledtabular}
\end{table}

\subsection{The transmission-phase integral
\texorpdfstring{$D_2=\zeta(3)-9/8$}{D2=zeta(3)-9/8}}
\label{app:phase}

The dispersive counterpart of $\kappa_2$ is fixed by the \emph{imaginary},
$\rho\times\rho$ leg combination of the self-energy~\eqref{eq:CdTg}---the
piece omitted in Eq.~\eqref{eq:legs} because it does not contribute to
$|T_{11}|^2$---which sets the transmission-phase coefficient $c_1$ of
Eq.~\eqref{eq:transphase}. Carrying out the same static-integral reduction
as for $D_2^{\kappa}$, but with both graviton legs in their regular
($\rho=r^3$) form, the frequency-independent content collapses to the
single weight-three constant
\begin{equation}
D_2=\int_1^\infty\!\left[\frac{Q_2(x)}{x+1}\,a(x)
   +\frac{P_2(x)}{x+1}\,b(x)\right]dx,
\label{eq:D2phase}
\end{equation}
with inner integrals (the $\rho\times\rho$ analogues of $A$ and $B$)
\begin{equation}
a(x)=\tfrac34(x-1)^2+\ln\tfrac{x+1}{2},\quad
b(x)=-\tfrac{21}{16}-H(x),
\end{equation}
and $H$ the dilogarithmic antiderivative of Eq.~\eqref{eq:Hfull}.

Unlike $D_2^{\kappa}$, this integral requires no integer-relation search.
Under the rationalizing substitution $u=2/(x+1)$ its integrand expands into
a finite, $\ln2$-free sum involving only four polygamma-closed master
integrals,
\begin{subequations}
\label{eq:phasemasters}
\begin{align}
\int_0^1\! u^{a}\,du&=\frac{1}{a+1},\\
\int_0^1\! u^{a}\ln u\,du&=-\frac{1}{(a+1)^2},\\
\int_0^1\! u^{a}\,\Li_2(u)\,du&=\frac{\zeta(2)}{a+1}-\frac{H_{a+1}}{(a+1)^2},\\
\int_0^1\! u^{a}\ln u\,\ln(1-u)\,du
   &=-\frac{\psi'(a+2)}{a+1}+\frac{H_{a+1}}{(a+1)^2},
\end{align}
\end{subequations}
the polygammas reduced to argument $1$ by the recurrence. Here $\psi'(z)$
denotes the trigamma function. The hard
Euler-sum master that obstructs a fully symbolic treatment of
$D_2^{\kappa}$ is absent here, and the integral evaluates in closed form
directly,
\begin{equation}
D_2=\zeta(3)-\frac{9}{8}=0.077056903\ldots,
\label{eq:D2zeta}
\end{equation}
verified symbolically. The transmission-phase coefficient of
Eq.~\eqref{eq:transphase} is then
\begin{equation}
c_1=2\pi^2\,D_2=2\pi^2\!\left(\zeta(3)-\tfrac98\right)=1.52104\ldots,
\end{equation}
in which only the overall prefactor $2\pi^2$ is a numerical identification
(to four digits; a first-principles derivation requires the dispersive
near/far-zone matching and is left to future work). The appearance of the
odd constant $\zeta(3)$---in contrast to the purely even-weight
$\kappa_2$---is the analytic expression of the reality structure noted
below Eq.~\eqref{eq:dTg}: the real $\rho\times\sigma$ leg feeds the
absorptive sector, the imaginary $\rho\times\rho$ leg the dispersive phase.

\section{Higher multipoles}
\label{app:higherl}

The construction of the main text is specific to the quadrupole only
in its final, closed-form step; the framework itself holds for every
$\ell\ge2$. We record here how the ingredients scale with $\ell$ and
give the numerical coefficients for the octupole, deferring
closed forms for $\ell>2$ to future work.

\subsection{General-\texorpdfstring{$\ell$}{l} structure}

The coupled system~\eqref{eq:coupled} and the symmetrized
potentials~\eqref{eq:potentials} are written for general $\ell$, with
the coupling strength entering through
\begin{equation}
\Voff=6\sqrt{\Lambda}\,\sqrt{\zzeta}\,\frac{M^3 f}{r^5},
\qquad \Lambda=\frac{(\ell+2)!}{(\ell-2)!},
\end{equation}
so the conversion vertex grows as $\sqrt{\Lambda}$
($6\sqrt{\Lambda}=12\sqrt6,\,12\sqrt{30},\ldots$ for
$\ell=2,3,\ldots$). The general-relativistic scalar-to-graviton
greybody ratio that normalizes the observables is
\begin{equation}
\begin{aligned}
\frac{\Gamma_{0,\ell}}{\Gamma_{2,\ell}}
&= \left[\frac{(\ell-2)!\,(\ell+2)!}{(\ell!)^2}\right]^{-2} \\[4pt]
&= \frac{1}{36},\;\frac{9}{100},\;\ldots \quad (\ell=2,3,\ldots).
\end{aligned}
\end{equation}
The scalar static solutions remain Legendre functions,
$p_\ell=r\,P_\ell(r-1)$ and $q_\ell=r\,Q_\ell(r-1)$, for every
$\ell$. What does \emph{not} carry over in elementary form is the
axial pair: the regular Regge--Wheeler static solution is the monomial
$\rho=r^3$ only for $\ell=2$, and for $\ell>2$ both the regular
solution and its irregular partner $\sigma_\ell$
[the analogue of Eq.~\eqref{eq:sigmaclosed}] must be constructed
anew. Consequently the conversion and self-energy integrals
$J_\ell$ and $D_\ell^{\kappa}$ are distinct for each multipole, and
their closed-form evaluation is a separate exercise of the type
carried out in Appendices~\ref{app:d2} and \ref{app:kappa}.

\subsection{Octupole coefficients}

Integrating the two-channel system numerically and extracting the
frequency-independent, $\zzeta$-linear
combinations~\eqref{eq:ddef}--\eqref{eq:kappadef} as in
Sec.~\ref{sec:checks}, we obtain for $\ell=3$
\begin{equation}
d_3=0.87(1),\qquad \kappa_3=0.85(1),
\end{equation}
to be compared with the closed-form quadrupole values
$d_2=0.283579$ and $\kappa_2=0.394775$
(Table~\ref{tab:multipole}). The larger $\ell=3$ coefficients suggest that the effect becomes more
pronounced, in normalized units, at higher multipole order, consistent
with the $\sqrt{\Lambda}$ growth of the vertex, though a single
additional multipole does not establish a general trend.
This growth is, however, offset in any physical observable by the
rapidly falling contribution of higher multipoles to the radiated
signal and by the steeper $(2M\omega)^{2\ell+2}$ suppression of their
greybody factors, so the quadrupole remains the dominant channel.

\begin{table}[htbp]
\caption{Parity-violating coefficients by multipole. The $\ell=2$
values are the exact closed form (Eq.~\eqref{eq:d2}) and the
high-precision reconstruction (Eq.~\eqref{eq:kappa2}); the $\ell=3$
values are numerical.}
\label{tab:multipole}
\begin{ruledtabular}
\begin{tabular}{cccc}
$\ell$ & $\Lambda$ & $d_\ell$ & $\kappa_\ell$ \\
\hline
$2$ & $24$  & $0.283579$ (exact) & $0.394775$ (reconstructed)\\
$3$ & $120$ & $0.87(1)$ & $0.85(1)$ \\
\end{tabular}
\end{ruledtabular}
\end{table}

A closed-form family $d_\ell,\kappa_\ell$ for general $\ell$---and the
associated spectral sums entering the total absorbed flux and scalar
luminosity---would require the general-$\ell$ axial static solutions
and is left to future work.

\section{The complete two-channel S-matrix at \texorpdfstring{$\ell=2$}{l=2}}
\label{app:smatrix}
 
The analysis of the main text follows a single incident channel---an axial
graviton impinging from infinity---and extracts the parity-violating
greybody coefficients $d_2$ and $\kappa_2$ from its reflected, transmitted,
and converted content.  For completeness we give here the full
infinity-incidence scattering matrix of the coupled axial-graviton
($i=1$) / pseudoscalar ($i=2$) system at $\ell=2$.  With a unit wave
incident from infinity in channel $j$, the coupled solution behaves as
\begin{align}
  \psi_i &\sim \delta_{ij}\,e^{-i\omega r_*}+R_{ij}\,e^{+i\omega r_*},
  & r_*&\to+\infty, \nonumber\\
  \psi_i &\sim T_{ij}\,e^{-i\omega r_*},
  & r_*&\to-\infty,
  \label{eq:smatrix-bc}
\end{align}
defining the $2\times2$ reflection and transmission matrices $R$ and $T$.
Both channels are massless, so the two asymptotic wavenumbers coincide and
flux conservation takes the channel-symmetric form
\begin{equation}
  R^\dagger R+T^\dagger T=\mathbb{1}.
  \label{eq:unitarity}
\end{equation}
The potential matrix is real and symmetric, so the full scattering matrix
is reciprocal, $S=S^{\mathsf T}$; in particular $R=R^{\mathsf T}$.  We solve
Eqs.~\eqref{eq:smatrix-bc} by the same DOP853 integration used throughout,
imposing purely ingoing data at the horizon in each channel in turn and
projecting onto the asymptotic basis at infinity.  The resulting
probabilities are collected in Table~\ref{tab:smatrix}; unitarity
\eqref{eq:unitarity} and reciprocity hold to $\sim10^{-12}$ across the
band, an end-to-end check of the coupled solver.
 
\begin{table}[htbp]
\centering
\caption{Two-channel S-matrix probabilities at $\ell=2$, $\zzeta=0.1$.
Channel $1$ is the axial graviton, channel $2$ the pseudoscalar:
$|T_{11}|^2$ is the axial greybody, $|T_{21}|^2$ the graviton$\to$scalar
conversion, $|T_{22}|^2$ the scalar greybody, and $|T_{12}|^2$ the
scalar$\to$graviton conversion.  The last column is the unitarity residual
$\big\|R^\dagger R+T^\dagger T-\mathbb{1}\big\|_{\max}$.  Both reflection
diagonals satisfy $|R_{ii}|^2=1-\Ord(10^{-3})$ and $R_{12}=R_{21}$ to
$\sim10^{-13}$.}
\label{tab:smatrix}
\begin{ruledtabular}
\begin{tabular}{ccccccc}
$M\omega$ & $|T_{11}|^2$ & $|T_{21}|^2$ & $|T_{22}|^2$ & $|T_{12}|^2$
 & residual \\
\colrule
$0.20$ & $9.17\times10^{-4}$ & $2.49\times10^{-5}$ & $2.34\times10^{-5}$ & $7.59\times10^{-6}$ & $4\times10^{-12}$\\
$0.15$ & $7.47\times10^{-5}$ & $1.99\times10^{-6}$ & $1.95\times10^{-6}$ & $5.97\times10^{-7}$ & $3\times10^{-13}$\\
$0.10$ & $3.28\times10^{-6}$ & $8.62\times10^{-8}$ & $8.72\times10^{-8}$ & $2.58\times10^{-8}$ & $2\times10^{-12}$\\
$0.07$ & $2.70\times10^{-7}$ & $7.04\times10^{-9}$ & $7.23\times10^{-9}$ & $2.11\times10^{-9}$ & $4\times10^{-13}$\\
$0.05$ & $2.90\times10^{-8}$ & $7.56\times10^{-10}$ & $7.81\times10^{-10}$ & $2.27\times10^{-10}$ & $2\times10^{-13}$\\
$0.03$ & $1.12\times10^{-9}$ & $2.92\times10^{-11}$ & $3.04\times10^{-11}$ & $8.80\times10^{-12}$ & $2\times10^{-13}$\\
\end{tabular}
\end{ruledtabular}
\end{table}
 
Three features of Table~\ref{tab:smatrix} bear on the main results.
First, the graviton-incidence column ($j=1$) reproduces the quantities of
the main text: the axial greybody $|T_{11}|^2$ supplies $\kappa_2$ through
its $\Ord(\zzeta)$ excess over the polar (GR) value, while the
graviton$\to$scalar conversion obeys
$|T_{21}|^2/(\zzeta\,\Gamma_{\rm pol})\to d_2=0.28358$ as $M\omega\to0$, in
agreement with the closed form $d_2=3(\pi^2-9)^2/8$.
 
Second, the pseudoscalar-incidence column ($j=2$) is new but
parity-even-dominated.  The scalar greybody $|T_{22}|^2$ is, at leading
order in $\zzeta$, the ordinary massless-scalar greybody of Schwarzschild
at $\ell=2$ and carries no parity-violating information; the \dCS{}
coupling enters it only at $\Ord(\zzeta)$.  The scalar$\to$graviton
conversion $|T_{12}|^2$ is a genuinely independent $\Ord(\zzeta)$
coefficient: it is \emph{not} equal to $|T_{21}|^2$, their ratio being
frequency-independent, $|T_{12}|/|T_{21}|\simeq0.548$.  The asymmetry is
physical---the axial graviton sees the shallower Regge--Wheeler barrier
($-6M/r^3$ versus $+2M/r^3$ for the scalar), penetrates more deeply into
the conversion region, and therefore converts more efficiently than the
scalar does in reverse.  (Reciprocity relates infinity-incidence to
horizon-incidence, $T^{\rm hor}=T^{\mathsf T}$, and does not constrain
$|T_{12}|$ against $|T_{21}|$.)
 
Third, and most important, the parity-violating observables are confined
to the graviton-incidence column: $\kappa_2$ resides in the axial diagonal
and $d_2$ in the off diagonal, and neither is modified by the
scalar-incidence sector.  The full S-matrix therefore \emph{completes} the
scattering problem without altering the coefficients reported in the main
text.
 
Finally, nothing in this structure is special to $\ell=2$.  For every
$\ell\ge2$ the S-matrix retains the same two-channel block form, the
diagonal greybodies follow the universal low-frequency law
$|T_{ii}|^2\propto(2M\omega)^{2\ell+2}$, and the conversion channels scale
as $\sqrt{\zzeta}$ in amplitude.  The parity-violating coefficients grow
mildly with multipole---e.g.\ $d_3\approx0.870$, $\kappa_3\approx0.850$
(Appendix~\ref{app:higherl})---but the effect is dominated by the
quadrupole, so we present the explicit matrix at $\ell=2$ and the general
$\ell$ follows by the same construction.

\bibliography{references}

\end{document}